\documentclass[aps,prl,superscriptaddress,twocolumn,nofootinbib,preprintnumbers]{revtex4-2}

\usepackage{amsmath}
\allowdisplaybreaks
\usepackage{graphicx}
\usepackage{subfigure}
\usepackage{amssymb}
\usepackage{booktabs}
\usepackage{xcolor}
\usepackage{multirow}
\usepackage{cancel}
\usepackage{color}
\usepackage{ulem}
\usepackage{listings}
\usepackage{xcolor}
\usepackage{float}
\usepackage{slashed}
\usepackage{wrapfig}
\usepackage{mathtools}
\usepackage[colorlinks,citecolor=blue]{hyperref}
\usepackage[T1]{fontenc}  
\usepackage[utf8]{inputenc}  
\usepackage{times}
\begin{document}
	
	\title{Dominant Thermal Resonant Mechanism for Low-Scale Leptogenesis}
	
	\author{Shao-Ping Li}
	\email{lisp@het.phys.sci.osaka-u.ac.jp}
	\affiliation{Department of Physics, The University of Osaka, Toyonaka, Osaka 560-0043, Japan}
    \affiliation{Marietta Blau Institute for Particle Physics, Austrian Academy of Sciences, Dominikanerbastei 16, A-1010 Vienna, Austria}
    
    \author{Apostolos Pilaftsis}
\email{apostolos.pilaftsis@manchester.ac.uk}
\affiliation{Department
 of Physics and Astronomy, University of Manchester, Manchester, M13 9PL, United Kingdom}

	\begin{abstract}
We explicitly demonstrate the importance of a new thermal resonant channel in the context of low-scale lepto\-genesis, which goes beyond the well-known mixing and oscillation of massive sterile neutrinos. This new channel is always present when considering the thermally-induced Higgs decay to leptons and relativistic\- sterile neutrinos, and can become dominant thanks to lepton-doublet flavour coherences being  resonantly enhanced in thermal plasma. This mechanism, which we call Thermally Induced Resonant Leptogenesis~(TIRL), can yield the observed baryon asymmetry in our universe. No resonant enhancement from quasi-degenerate sterile neutrinos is required, providing  distinction between TIRL  and other known low-scale leptogenesis scenarios. The active-to-sterile neutrino mixing can be also probed in fixed-target and long-lived particle experiments and by displaced vertex searches at high-energy colliders.
	\end{abstract}
	
\maketitle
\preprint{OU-HET 1299} 
\textbf{\textit{Introduction}}. Low-scale leptogenesis may successfully be realised within the type-I seesaw extension of the Standard Model (SM), with the latter being studied extensively because of its simple solutions to the neutrino mass origin~\cite{Minkowski:1977sc,Yanagida:1980xy,Mohapatra:1979ia,Schechter:1980gr} and the Baryon Asymmetry of the Universe (BAU)~\cite{Fukugita:1986hr}. Low-scale leptogenesis utilizes electroweak or GeV-scale heavy neutrinos, which can give rise to various detectable signatures in collider and laboratory experiments~\cite{Atre:2009rg,Deppisch:2015qwa,Abdullahi:2022jlv}, such as the fixed-target experiments NA62~\cite{NA62:2017rwk} and SHiP~\cite{Alekhin:2015byh,SHiP:2018xqw}, long-lived particle experiments FASER~\cite{FASER:2018eoc} and MATHUSLA~\cite{Chou:2016lxi,Curtin:2018mvb},   the long baseline oscillation experiment DUNE~\cite{DUNE:2018tke}, displaced lepton jet or displaced vertex search at LHC and high-luminosity LHC~\cite{Izaguirre:2015pga,Cottin:2018nms,Drewes:2019fou}, as well as future hadron and lepton colliders~\cite{Datta:1993nm,Pilaftsis:1991ug,Kersten:2007vk,Dev:2013wba,Antusch:2016ejd,Antusch:2017pkq,Harz:2021psp}.

Low-scale leptogenesis resonantly enhanced due to quasi-degenerate heavy neutrino mixing, known as resonant lepto\-genesis~(RL)~\cite{Pilaftsis:1997jf,Pilaftsis:2003gt,Pilaftsis:2005rv}, and via sterile neutrino oscillations (the ARS mechanism)~\cite{Akhmedov:1998qx,Asaka:2005pn} long stand out as being the  leading paradigms that have attracted much attention in the literature, 
e.g.~see~Ref.~\cite{Davidson:2008bu} for a review. In general, certain degree of degeneracy exists in the sterile neutrino mass spectrum both in the ARS mechanism and RL, to enhance the BAU.  As such,  the parameter space is significantly dependent on the degree of degeneracy. 

In this Letter, we present a new and potentially dominant mechanism  for low-scale leptogenesis within the type-I seesaw framework, which differs from RL and the ARS scenario.  This new channel is always present  at finite temperatures when the SM Higgs $\Phi$ can decay to lepton doublets $L_\alpha$ and relativistic sterile neutrinos $N_i$, owing to thermal mass effects. 
While thermal Higgs decay to GeV-scale sterile neutrinos has been noted earlier~\cite{Pilaftsis:1997jf,Giudice:2003jh},  explicit realization of  this channel was  found to require a high degree of mass degeneracy~\cite{Hambye:2016sby} and hence still relies on the traditional RL mechanism. Contrary to this well-studied scenario, we show that the BAU problem can be solved  from  SM Higgs decay without requiring any mass degeneracy.  Remarkably, we find a significant enhancement for leptogenesis induced by \textit{resonant thermal-lepton coherences}, which  is not achieved by suitably arranging the unknown parameters of the type-I seesaw framework, but is mainly specified by the SM sector. 

\begin{figure}[t]
	\centering
\includegraphics[scale=0.6]{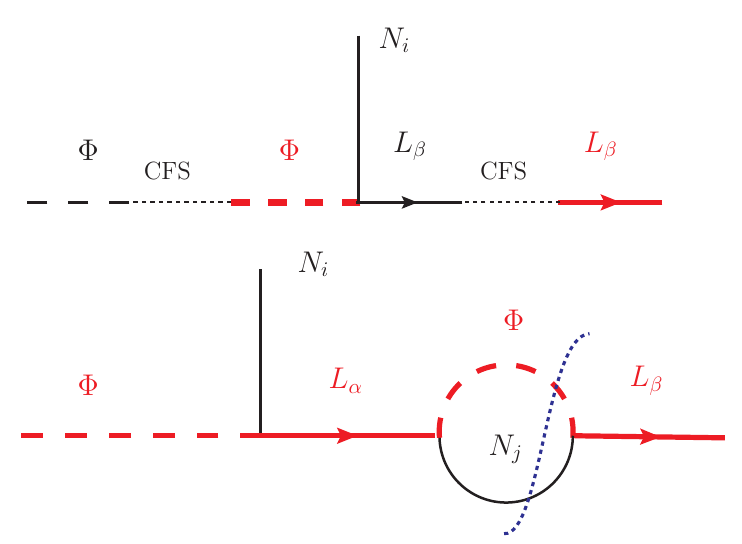}
	\caption{\label{fig:Hdec}
    Thermally-induced tree-level and one-loop decay of the Higgs doublet~$\Phi$ into lepton doublets $L_\alpha$ and light sterile neutrinos $N_i$, where the red, thick lines represent thermal resummation effects via Coherent Forward Scattering of $\Phi$ and $L_{\alpha,\beta}$ with the background plasma. The absorptive part of the loop decay amplitude induced by the thermal cut is indicated by the blue-dotted line.
    }
\end{figure}

By deriving the pertinent transport equations within a flavour-covariant density matrix formalism aided by non-equilibrium~quantum field theory~\cite{Prokopec:2003pj,Prokopec:2004ic,Beneke:2010dz,BhupalDev:2014pfm,BhupalDev:2014oar}, we demonstrate how the thermal lepton-doublet oscillation source can boost the generation of lepton asymmetry in the thermally allowed Higgs decay\- $\Phi \to L_\alpha N_i$, shown in Fig.~\ref{fig:Hdec}. We call this mechanism Thermally Induced Resonant Lepto\-genesis~(TIRL).  The resulting parameter space in TIRL is predictive because of a well-determined enhancement factor, 
and the viable parameter space with two GeV-scale sterile neutrinos  can be  explored by MATHUSLA,  SHiP,  current hadron collider LHC, and future lepton colliders such as CEPC~\cite{CEPCStudyGroup:2018ghi} and FCC-ee~\cite{FCC:2018evy}.

\medskip
\textbf{\textit{Flavour-covariant transport equations}}.
To fix our notation, let us first recall the relevant Yukawa Lagrangian, 
\begin{equation}
   \label{eq:LY}
-\,\mathcal{L}_{\rm Y}\ =\: y_{\alpha\beta'}\bar L_\alpha \Phi P_R E_{\beta'}\: +\: h_{\alpha j}\bar L_\alpha \widetilde{\Phi}P_R N_{j}\: +\: \rm H.c., 
\end{equation}
where $P_R=(1+\gamma_5)/2$ is the right-handed chirality projector, $\Phi$ is the SM scalar doublet,  $\widetilde{\Phi}=\text{i} \sigma_2 \Phi^*$ is its hyper\-charge conjugate,  and $L_\alpha, E_{\alpha'}, N_i$ denote lepton doublets, right-handed charged leptons and right-handed singlet neutrinos, respectively. The latter particle species can carry mass terms that can be either Majorana type, ${M^{\rm M}_{ij}\bar{N}_i^C P_R\, N_j + {\rm H.c}}$, or Dirac type, ${M^{\rm D} _{ij}\bar{N}_i P_R N_j + {\rm H.c.}}$. If the singlet Majorana neutrino masses are much smaller than the temperatures during which leptogenesis  is mostly active,  they can populate in the relativistic regime where the helicity acts as an approximately  conserved quantum number.
The generalised lepton number $L_{\rm tot}=L_{\rm SM}+L_N$~\cite{Akhmedov:1998qx,Asaka:2005pn,Abada:2018oly} can then be defined, where $L_{\rm SM} = L_L + L_E$ and $L_N$ denote the SM lepton number and $N$-sector helicity number, respectively. In this Letter, we consider relativistic Dirac sterile neutrinos that conserve $L_{\rm tot}$ to build the quantum transport equations, but our results will still hold for singlet Majorana neutrinos in the type-I seesaw framework to a good approximation, as long as the $L_{\rm tot}$-violating rates are suppressed by~${M^2/T^2\ll 1}$.

To effectively describe the new channel in TIRL, we write down the relevant flavour-covariant transport equations in the density-matrix formalism~\cite{Sigl:1993ctk,BhupalDev:2014pfm}
\begin{eqnarray}
    \label{eq:ave-nL}
 \dot{\underline{n}}^L \!\!&\simeq&\!\! -\, \frac{i}{2}\,\Big[\,{\cal E}_L\,, \Delta n^L\Big]\: -\: \frac{1}{2n^N_{\rm eq}n^L_{\rm eq}}\,
\Big\{ \underline{n}^L\,, \text{Tr}_{\bar{N}}\big(\underline{n}^N \underline{\gamma}^{\bar{N}L}_{\bar{\Phi}} \big)\Big\}\: 
\nonumber\\
\!\!&&\!\!+\:  \text{Tr}_{\bar{N}}\big(\underline{\gamma}^{\bar{\Phi}}_{\bar{N}L}\big)\ +\ \dots\,,\\[3mm] 
   \label{eq:DnL}
\Delta \dot{n}^L \!\!&\simeq&\!\! -\, 2i\,\Big[\,{\cal E}_L\,, \underline{n}^L\Big]\: 
-\: \frac{1}{2n^N_{\rm eq}n^L_{\rm eq}}\,
\Big\{ \underline{n}^L\,, \text{Tr}_{\bar{N}}\big(\underline{n}^N \delta\gamma^{\bar{N}L}_{\bar{\Phi}} \big)\Big\}\nonumber\\
\!\!&&\!\!-\: \frac{1}{2n^N_{\rm eq}n^L_{\rm eq}}\,
\Big\{ \Delta n^L\,, \text{Tr}_{\bar{N}}\big(\underline{n}^N \underline{\gamma}^{\bar{N}L}_{\bar{\Phi}} \big)\Big\}\\ 
\!\!&&\!\!+\: \frac{1}{2n^N_{\rm eq}n^L_{\rm eq}}\,
\Big\{ \underline{n}^L\,, \text{Tr}_{\bar{N}}\big(\Delta n^N \underline{\gamma}^{\bar{N}L}_{\bar{\Phi}} \big)\Big\}
-\text{Tr}_{\bar{N}}\big(\delta\gamma^{\bar{\Phi}}_{\bar{N}L}\big)\,.\nonumber
\end{eqnarray}
In~\eqref{eq:ave-nL} and~\eqref{eq:DnL}, the dot denotes differentiation with respect to cosmic time~$t$, ${\cal E}_L$ is the thermally averaged self-energy matrix, $\underline{n}^X\!\equiv\! \frac{1}{2}(n^X + \bar{n}^X)$ is the CP$\widetilde{\rm T}$-even number density of $X = N_i, L_\alpha$ spieces, and ${\Delta n^X \equiv n^X - \bar{n}^X}$ is their respective CP$\widetilde{\rm T}$-odd number density, where $\widetilde{\text{T}}$ denotes {\it naive} time-reversal transformation~\cite{BhupalDev:2014pfm}. Similarly, ${\underline{\gamma}^X_{YZ} \equiv \frac{1}2\big[\gamma^{X}_{YZ} + (\gamma^{\bar{X}}_{\bar{Y}\bar{Z}})^{\sf T}\big]}$ and ${\delta \gamma^X_{YZ} \equiv \gamma^{X}_{YZ} -(\gamma^{\bar{X}}_{\bar{Y}\bar{Z}}})^{\sf T}$ are the flavour-covariant 
CP$\widetilde{\rm T}$-even and CP$\widetilde{\rm T}$-odd collision terms, with $X,Y,Z \in \{N_i, L_\alpha, \Phi\}$, 
where transposition $\sf T$ acts on both the $N$ and $L$ flavour spaces. In addition,
$\text{Tr}_X(\cdots )$ is a trace taken {\it only} over the $X$ flavour space. 
Finally, we ignore subdominant contributions to the left-handed leptonic asymmetry $\text{Tr}_L\Delta n^L$ from right-handed charged leptons $E_{\alpha'}$ and SM gauge 
bosons. A thermally complete description of the entire $N,L,E$ system is presented in the Supplemental Material~\cite{SM}.

The ellipsis on the right-hand side of~\eqref{eq:ave-nL} indicates the omission of equilibrated right-handed charged lepton Yukawa and gauge interactions. These effects lead to decoherence in the left-handed lepton doublets such that $\underline{n}^L\simeq n^L_{\rm eq}$ with $\dot{\underline{n}}^L\simeq 0$
is a good approximation. Then, the 
commutator $\big[\,{\cal E}_L\,, \Delta n^L\big]$ in~\eqref{eq:ave-nL}, usually ignored in RL and ARS scenarios, becomes the dominant source of $L_\alpha$-$L_\beta$ transitions in TIRL, which are induced by thermal mass differences,~$\tilde m_\alpha^2-\tilde m_\beta^2 =(y_\alpha^2-y_\beta^2)T^2/16$ (with $y_{\alpha\alpha'} \equiv y_{\alpha}\delta_{\alpha\alpha'}$), rather than vacuum mass differences. Crucially, as we will see below, $\big[\,{\cal E}_L\,, \Delta n^L\big]$ will contribute to the final lepton asymmetry $\Delta n^L_{\alpha\alpha}$ through the second line in~\eqref{eq:DnL}. Such a contribution to $\Delta n^L_{\alpha\alpha}$ consistently occurs in a proper flavour-covariant formulation of the lepton transport equations, while its dramatic impact has not been considered before in RL and ARS scenarios.

Since $\dot{\underline{n}}^L\simeq 0$ and $\underline{n}^L_{\alpha\beta}\simeq (n^L_{\rm eq})_{\alpha\beta}\,\delta_{\alpha\beta}$ in the basis $y_{\alpha\alpha'} \equiv y_{\alpha}\delta_{\alpha\alpha'}$, \eqref{eq:ave-nL} becomes a constraint equation for the off-diagonal coherences $\Delta n^L_{\alpha\neq\beta}$. 
From~\eqref{eq:ave-nL}, we then find that
\begin{align}\label{eq:Dnab}
   \Delta n^L_{\alpha\ne \beta}\: \simeq\: 2i\, h_{\alpha j}h^*_{\beta j}\,\frac{T\,\Delta m_j^2}{\tilde{m}^2_\alpha-\tilde{m}^2_\beta}\int_{{\bf k}_{L,\Phi,N}}\!\!\!\tilde{\delta}^{(4)}(k)\,\mathcal{F}_j\,.
\end{align}
Here, $\int_{{\bf k}_X}\equiv \int {\rm d}^3 {\bf k}_X/((2\pi)^3 2E_X)$ is the integral phase-space  measure of the $X$ particle, $\tilde{\delta}^{(4)}(k)\equiv (2\pi)^4 \delta^{(3)}({\bf k}_L +{\bf k}_\phi-{\bf k}_N)\delta(E_L +E_N-E_\phi)$,   $\Delta m^2_j\equiv (M_\Phi^2-\tilde{m}^2_L - m_{N_j}^2)^{1/2}$, where $M_\Phi^2$ includes the vacuum and thermal mass of the SM scalar~\cite{Carrington:1991hz}, and  $\mathcal{F}_j\equiv \delta f^N_{j}(E_N) \big(f^\Phi_{\rm eq}(E_\Phi)+f^L_{\rm eq}(E_L)\big)$ with $\delta f^N\equiv f^N-f^N_{\rm eq}<0$. We have adopted the occupation-number matrices for $L_\alpha$ and $\bar{L}_{\alpha}$, $f^L_{\alpha\beta}(k_0)=[e^{(k_0-\mu)/T}+1]_{\alpha\beta}^{-1}$ and $\bar f^L_{\alpha\beta}(-k_0)=[e^{(-k_0+\mu)/T}+1]_{\alpha\beta}^{-1}$, where the effective chemical potentials $\mu_{\alpha\beta}$ characterise a small lepton-number asymmetry created by out-of-equilibrium processes, mediated by the small neutrino Yukawa couplings $h_{\alpha j}$ in~\eqref{eq:LY}. The appearance of the thermal mass difference $\tilde m_\alpha^2-\tilde m_\beta^2$ in~\eqref{eq:Dnab} is induced by $\big[\,{\cal E}_L\,, \Delta n^L\big]$, which, due to the small charged-lepton Yukawa couplings, leads to a significant enhancement of the lepton-flavour coherences. 
An equivalent calculation of~\eqref{eq:Dnab} using the flavour-covariant Kadanoff-Baym formulation~\cite{Prokopec:2003pj,Prokopec:2004ic,Beneke:2010dz,BhupalDev:2014pfm,BhupalDev:2014oar} is presented in a companion paper~\cite{Li:2026qym},  demonstrating that~\eqref{eq:Dnab} is consistent with the lepton-number conservation $\Delta L_L+\Delta L_N=0$.

To derive the observable CP-odd (CP$\widetilde{\rm T}$-odd) leptonic asymmetry $\Delta n^L_{\alpha\alpha}$ for a given lepton doublet  $L_\alpha$, we substitute $\Delta n^L_{\alpha\ne \beta}$ from~\eqref{eq:Dnab} into~\eqref{eq:DnL}. Since the lepton-flavour coherences are already at $\mathcal{O}(h^2)$,
this will produce a CP-violating effect at $\mathcal{O}(h^4)$ from the nonthermal neutrino Yukawa interactions, i.e., the second line on the right-hand side of~\eqref{eq:DnL}. 
Further, the CP-violating effect is enhanced by the factor  $\propto T/(\tilde m^2_\alpha-\tilde m^2_\beta)$ in~\eqref{eq:Dnab}. Thus,  one  expects that the smallness of the charged-lepton couplings~$y_\alpha$  would create an enhancement large enough to compensate for the smallness of the couplings~$h_{\alpha j}$  in freeze-in leptogenesis, without resorting to highly degenerate signlet neutrinos, as done in~\cite{Hambye:2016sby,Hambye:2017elz}. The resonant enhancement of the BAU due to the thermally induced $L_\alpha$-$L_\beta$ coherent transitions justifies the name TIRL.  It is important to note that the RL in~\cite{Hambye:2016sby,Hambye:2017elz} is a lepton-number violating effect of $\mathcal{O}(h^4)$ due to Majorana masses, ARS is an $\mathcal{O}(h^6)$ lepton-number conserving effect, while TIRL is an $\mathcal{O}(h^4)$, lepton-number conserving effect. These are the main differences among the three low-scale leptogenesis models.

\textbf{\textit{TIRL from lepton-doublet flavour coherences}}. We may rewrite the evolution of the lepton-doublet flavour asymmetry\- $Y_\alpha\equiv \Delta n^L_{\alpha\alpha}/s$ in the compact form,
\begin{align}
   \label{eq:Yalpha}
    sH z\frac{\text{d}Y_\alpha}{\text{d}z}\: =\: \mathcal{S}_\alpha+\mathcal{W}_\alpha \,,
\end{align}
where $H\approx 2.2\times 10^{-14}/z^2$~GeV, $s\approx 9.1\times 10^7/z^3~\text{GeV}^3$ denote the Hubble parameter and the SM entropy density, respectively, and $z\equiv m_h/T$ with $m_h\approx 125$~GeV the vacuum mass of the SM scalar. The lepton-number asymmetry stored in lepton doublets will   be converted into baryon asymmetry via $Y_B=-2/3\times \sum_\alpha Y_\alpha$~\cite{Harvey:1990qw}, which should match the observed value today $Y_B\simeq 8.75\times 10^{-11}$~\cite{Planck:2018vyg}. 

We calculate the washout rate $\mathcal{W}_\alpha$ from Higgs decay and its inverse decay $\bar{\Phi} \leftrightarrow L \bar{N}$ in~\eqref{eq:DnL}, while neglecting spectator effects that redistribute the asymmetry of lepton doublets among all thermalised SM particles. In this way, we find  
\begin{align}
    \mathcal{W}_\alpha\approx -6.2\times 10^7|h_\alpha|^2z^{-4}Y_
\alpha~\text{GeV}^4\,,
\end{align}
where $|h_\alpha|^2\equiv \sum_i  |h_{\alpha i}|^2$. The dominant contribution to the source term $\mathcal{S}_\alpha$ in 
\eqref{eq:Yalpha} originates from the third collision term on the right-hand side of~\eqref{eq:DnL}. After including quantum statistical factors for the fields $L_\alpha$ and $\bar{N}_i$, we explicitly obtain
\begin{align}\label{eq:Salpha}
    \mathcal{S}_\alpha= T^2\!\!\sum_{i\ne j,\beta\ne \alpha}\!\!\frac{\Delta m_i^2\Delta m_j^2\text{Im}(h^*_{\alpha i}h_{\alpha j}h_{\beta i}h^*_{\beta j})}{16\pi^4\,(\tilde{m}^2_\beta-\tilde{m}^2_\alpha)}\,\langle\delta f^{N}_{ij}\rangle\,,
\end{align}
where the dimensionless quantity $\langle\delta f^{N}_{ij}\rangle$ reads
\begin{align}
    \langle\delta f^{N}_{ij}\rangle=& \int_0^\infty \frac{\text{d}x_\ell}{x_\ell}\int_{\tilde x_N}^\infty \text{d}x_N\text{d}x_N' I^{\ell}_{\rm eq} I^{N}_i \delta f^{N}_j(x_N')\,.
\end{align}
Here, $\tilde x_N\approx 0.1/x_\ell$ for both $x_N$ and $x_N'$ integration, with  $x^{(\prime)}_i\equiv |{\bf p}^{(\prime)}_i|/T$,  $I^{\ell}_{\rm eq}\equiv f^\Phi_{\rm eq}(x_\ell+x_N') + f^L_{\rm eq}(x_\ell)$ and ${I^{N}_i\equiv f^\Phi_{\rm eq}(x_\ell+x_N)+f^{N}_i(x_N)}$. The  CP-violating source $\mathcal{S}_\alpha$  depends on the nonthermal neutrino distribution function that should be determined numerically. For this, we solve the Boltzmann equations for $f^N$ by including the Higgs decay, inverse decay, as well as the corrections from soft gauge-boson interactions~\cite{Anisimov:2010gy,Besak:2012qm,Garbrecht:2013bia,Laine:2013lka}.  Note that if the dependence of $\langle\delta f^{N}_{ij}\rangle$ on the $i$-flavour is weak, the dependence on the $j$-flavour must be strong. Otherwise, the summation over $i,j$ would give  $\text{Im}[(h h^{\dagger})_{\beta\alpha}(h h^\dagger)_{\alpha\beta}]=0$.

To numerically illustrate the importance of TIRL at low scales, 
we adopt the Casas-Ibarra  parametrization~\cite{Casas:2001sr}: $h_{\alpha i}= i \sqrt{2}\sum_{\beta}U_{\alpha \beta}\sqrt{m_\beta}\mathbb{R}_{\beta i}\sqrt{M_i}/v$ to perform the numerical evaluation,  where $v\approx 246$~GeV is the electroweak vacuum expectation value, $U$ is the Pontecorvo-Maki-Nakagawa-Sakata (PMNS) in the standard parametrization of~\cite{ParticleDataGroup:2024cfk}, $m_\beta$ with $\alpha=1,2,3$ are the SM neutrino masses, and $\mathbb{R}$ is a complex orthogonal matrix.
A characteristic feature of the CP-violating source is that $\mathcal{S}_{\alpha}$ follows the strong hierarchy of the lepton masses within the SM.  In particular, the largest contribution occurs in the muon-electron flavour sector, with $\mathcal{S}_\mu(\alpha=2,\beta=1)=\mathcal{S}_e(\alpha=1,\beta=2)$. In evaluating the lepton asymmetry, we therefore take this largest CP-violating source for consideration. In addition, we will only take two neutrino generations into account by decoupling the third neutrino singlet, which we specify as $N_3$. Although three sterile neutrinos generally open a more viable parameter space~\cite{Abada:2018oly,Drewes:2021nqr}, $N_3$ is restricted to be at the same mass scale as $N_1, N_2$. However, our treatment leaves room for $N_3$ to create  additional effects in astrophysical and cosmological probes.  In particular, a light $N_3$ much below the GeV scale can have a longer lifetime to yield new observational imprints  on the spectral distortions of the cosmic microwave background by injecting extra  neutrino or photon energy~\cite{Chluba:2016bvg,Li:2025clq}, or  may contribute to the dark matter relic density and  astrophysical signatures~\cite{Dodelson:1993je,Shi:1998km,Asaka:2005pn,Asaka:2005an,Bezrukov:2025ttd}, which allows more complementary experimental signatures to those predicted by the BAU explanation  via two GeV-scale sterile neutrinos.  Therefore, we decouple  $N_3$ via $M_3\ll M_1, M_2$. Since $h_{\alpha i}\propto \sqrt{M_i}$ under Casas-Ibarra parametrization, the associated Yukawa couplings $h_{\alpha 3}$ are much smaller than $h_{\alpha 2}, h_{\alpha 1}$, ensuring that only two sterile neutrinos contribute significantly to the BAU. Nevertheless, making the contribution of $N_3$ to the BAU sufficiently small does not necessarily require that the lightest SM neutrino be massless, as can be inferred from the seesaw relation $m_{\nu,\alpha\beta}=-v^2h_{\alpha i}h_{\beta i}/M_i$, in line with the Casas-Ibarra parametrization.

We adopt the central values of the PMNS matrix mixing angles and neutrino mass-squared differences from NuFit-6.1~\cite{Esteban:2024eli}, and consider $M_1,M_2$ in the range of $[1,50]$~GeV. We develop a routine for parameter scan from an unbiased Markov Chain Monte Carlo (MCMC) to a biased one. The  biased data is then trained by Deep Neural Network (DNN) following a normalizing flow based on the real-valued non-volume preserving transformations~\cite{Dinh:2016q}. More details of the numerical routine are documented in the Supplemental Material~\cite{SM}.
For an unbiased MCMC scan, $10^6$ sample points are taken, with the aim of obtaining the correct order $Y_B=[10^{-11},10^{-10}]$. We find that a ratio of the Yukawa couplings at $0.01<|h_e|/|h_\mu|<0.1$ is favoured in the normal ordering of active neutrino mass spectrum, while $0.01<|h_\mu|/|h_e|<0.1$  is instead preferred in the inverted ordering. Here $|h_e|/|h_\mu|=\sqrt{|h_e|^2/|h_\mu|^2}$  characterises the relative washout effect between the electron and muon flavour asymmetries. The different hierarchical washout patterns in the normal and inverted ordering motivate us to conduct an efficient biased MCMC sampling by imposing the hierarchy prior $|h_\mu|/|h_e|>10$ for the normal ordering and  $|h_e|/|h_\mu|>10$ for the inverted ordering. The former corresponds to a scenario of muon dominance, where the total interaction rate of sterile neutrinos is dominated by couplings to the muon, whereas the latter reflects a scenario of electron dominance.  We then project the parameter space from the DNN-trained data onto the plane of $M_1$ and the active-sterile mixing angle $\theta_\alpha^2\equiv \sum_{i} v^2|h_{\alpha i}|^2/M^2_i$, in a slice of the normalized mass difference $\Delta M/M_1$ with $\Delta M\equiv M_2-M_1$. Such projections allow us to assess the reach of GeV-scale neutrino searches in experiments and for a better comparison with the sterile-neutrino mass spectrum in RL and ARS models.  

\begin{figure}[t]
	\centering
\includegraphics[width=\columnwidth]{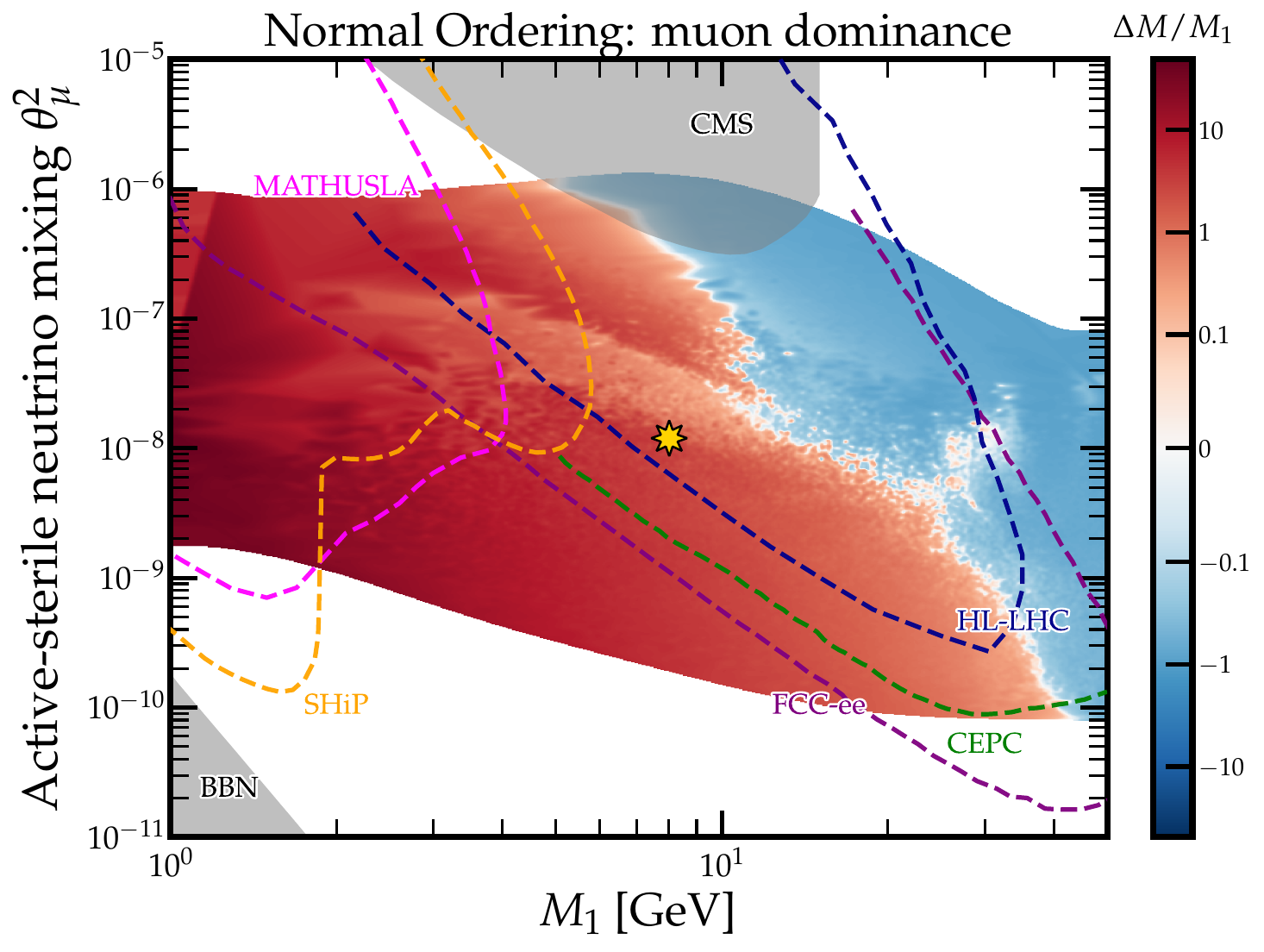} 
\\[0.2cm]
\includegraphics[width=\columnwidth]{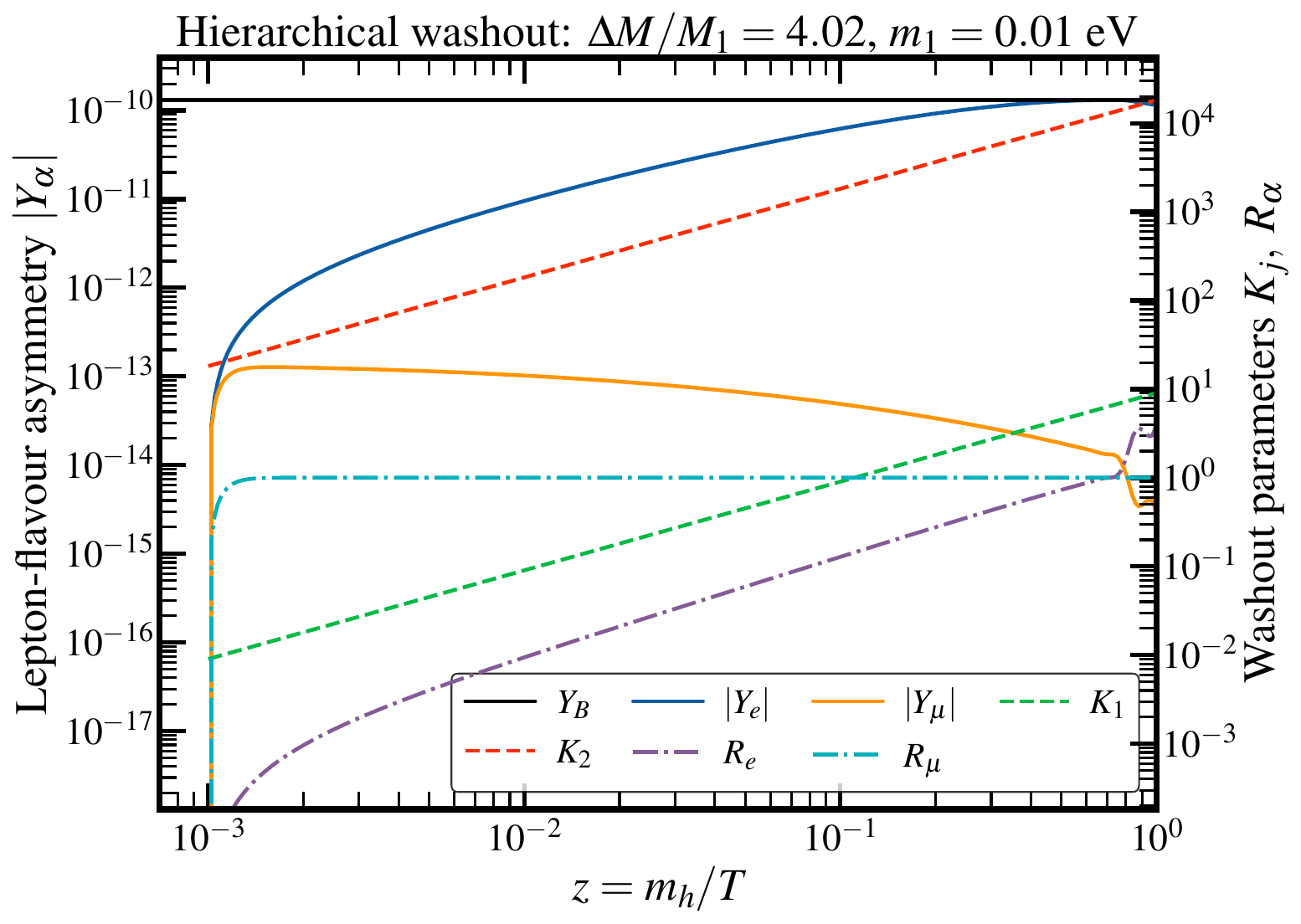}
	\caption{\label{fig:NO_Lm1} The parameter space of TIRL obtained from the scan routine: unbiased MCMC$\to$biased MCMC$\to$DNN training. \textit{Upper panel}: muon dominance preferred in normal ordering, with $\theta_\mu^2\gg \theta_e^2$,  $m_1=0.01$~eV for the lightest active neutrino mass and $\Delta M\equiv M_2-M_1$, yielding $Y_B=[10^{-11},10^{-10}]$. 
    The gray shaded regions are excluded by CMS and BBN.  The dashed lines represent the projected detection~limits. \textit{Lower panel}: a benchmark scenario (the yellow star from the upper) showing how the evolution of $|Y_e|=-Y_e,|Y_\mu|=-Y_\mu$ reaches the observed $Y_B$, and the hierarchical washout effects parametrized by $K_j$ and $R_\alpha$, with $M_1=8.0$~GeV and $M_2=40.2$~GeV. See text for more details.}
\end{figure}

Figure~\ref{fig:NO_Lm1} shows the result for  the normal ordering with a moderate mass $m_1=0.01$~eV for the lightest active neutrino. 
 For the inverted ordering,  except for the different flavour washout patterns,  the parameter space  is basically similar to the normal ordering~\cite{SM}.
 The upper panel explains the general parameter space for the correct order of $Y_B$, together with the current bounds from CMS~\cite{CMS:2022fut} and big-bang nucleosynthesis (BBN)~\cite{Boyarsky:2020dzc}.  The projected detection sensitivities from MATHUSLA~\cite{Curtin:2018mvb}, SHiP~\cite{SHiP:2018xqw}, displaced vertex searches at HL-LHC~\cite{Drewes:2019fou}, future lepton colliders CEPC~\cite{Antusch:2017pkq} and  FCC-ee~\cite{Blondel:2014bra} are also shown. From the upper panel, we also see that a mass hierarchy  between $M_1$ and $M_2$ is generically needed near 1~GeV and 50~GeV. When $M_1$ is around  1~GeV, $M_2$ should be much larger than $M_1$ with $M_2/M_1=\mathcal{O}(10)$.  This is understood by the fact that when both $N_1$ and $N_2$ are light near 1~GeV, the Yukawa couplings $h_{\alpha 1}, h_{\alpha 2}$ will be small. Since the enhancement of $\mathcal{S}_\mu=\mathcal{S}_e$ in TIRL does not rely on the adjustment of the mass difference $M_1-M_2$ but is rather fixed by the smallness of $y_\mu,y_e$, the dependence of the final lepton asymmetry on the sterile-neutrino masses is mainly through $h_{\alpha 1}\propto \sqrt{M_1},h_{\alpha 2}\propto \sqrt{M_2}$. On the other hand, when $M_1$ increases towards 50~GeV, $M_2$ becomes smaller than $M_1$. Nevertheless, there is no definite mass hierarchy between $N_1$ and $N_2$ around $10$~GeV. The different mass spectrum in sterile neutrinos also allows a direct comparison of the  general parameter space between TIRL  and ARS or RL. In particular, TIRL features $M_2/M_1\gg 1$ at $M_1\sim 1$~GeV and  $|\Delta M|\sim M_1$ at $M_1\sim 50$~GeV  in the upper boundary, where $\sum_\alpha\theta_\alpha^2\approx\theta_\mu^2\simeq 10^{-7}\text{--}10^{-6}$. However, as can be seen in~\cite{Klaric:2020phc,Klaric:2021cpi},   in ARS nearly half of the same region becomes unavailable to explain the BAU and the other half requires a mass splitting $\Delta M/M<10^{-6}$. Remarkably, such a tiny mass splitting is also the threshold when the Higgs decay\- $\Phi \to L_\alpha N_i$ via RL becomes dominant over TIRL~\cite{Hambye:2016sby,Hambye:2017elz,Li:2026qym}. Despite this, TIRL has distinctive features in comparison to the ARS and RL mechanisms. For instance, given that a mass splitting  $\Delta M \gg 1~\mathrm{GeV}$  is possible in TIRL, which is much larger than the typical invariant mass resolution for GeV-scale displaced decays at present and upcoming colliders~\cite{CMS:2022fut,ATLAS:2022atq}, the invariant mass of the sterile-neutrino decay products can in principle be reconstructed separately for each state, leading to two resolvable resonance peaks in the invariant mass distribution. However, for $\Delta M/M\lesssim 10^{-6}$, it becomes harder, if possible, to resolve the two states as separate mass peaks.

There is also a noticeable difference between the quasi-degenerate and non-degenerate sterile-neutrino mass spectra: the scaling of the total mixing angle $\sum_\alpha\theta^2_\alpha$ in terms of sterile neutrino masses. In the quasi-degenerate case $M_1\approx M_2\equiv M_N$, the seesaw relation $m_{\nu,\alpha\alpha}=-v^2 h_{\alpha 1}^2/M_1-v^2 h_{\alpha 2}^2/M_2\approx -\sum_iv^2 h_{\alpha i}^2/M_N$ implies a  simple relation $\sum_\alpha\theta^2_\alpha\sim \sum_\alpha m_{\nu,\alpha\alpha}/M_N\propto 1/M_N$. Indeed, this is observed to a good approximation in ARS~\cite{Canetti:2012kh,Klaric:2020phc}.  However, since TIRL does not depend on the quasi-degeneracy of $M_1, M_2$, a significant mass splitting between $N_1$ and $N_2$ will break down the simple scaling.  The boundaries of the parameter space shown in the upper panel of Fig.~\ref{fig:NO_Lm1} confirm this expectation, where the muon-dominated mixing  $\sum_\alpha\theta^2_\alpha\approx \theta^2_\mu\approx |h_{\mu 1}|^2v^2/M_1^2+|h_{\mu 2}|^2v^2/M_2^2$ depends on the combination of $1/M_1$ and $1/M_2$.

The lower panel of Fig.~\ref{fig:NO_Lm1} shows a benchmark scenario--the yellow star from the upper panel--for the evolution of electron and muon flavour asymmetries. The benchmark scenario explains  the evolution of different lepton flavour asymmetries in hierarchical washout.  To characterize the washout patterns, we consider the conventional parameter $K_j\equiv \Gamma_j/H$ defined as the production (thermalization) rate of sterile neutrinos  over the Hubble parameter, with  $\Gamma_j\approx 0.01 |h_{j}|^2 T$~\cite{Anisimov:2010gy,Besak:2012qm,Garbrecht:2013bia,Laine:2013lka} and $|h_j|^2\equiv \sum_\alpha |h_{\alpha j}|^2$.  We also consider the ratio of the washout to the source term $R_{\alpha}\equiv |\mathcal{W}_\alpha|/|\mathcal{S}_\alpha|$. The parameter $K_j$ measures whether and when $N_j$  becomes thermally populated in leptogenesis, while the ratio $R_\alpha$  signals if the washout dominates the evolution of $Y_{\alpha}$ and hence directly determines which flavour asymmetry survives. The muon dominance observed in normal ordering  indicates that the washout effect in $Y_\mu$ is much stronger, so that $Y_e$ eventually survives, which  
is guaranteed by the fact that $R_e>1$ only shows up at the very end of leptogenesis $z=m_h/T_{\rm sph}\approx 0.95$ with $T_{\rm sph}\approx 132$~GeV the sphaleron decoupling temperature~\cite{DOnofrio:2014rug}. In addition, $K_2\gg 1$ indicates that $N_2$ comes into equilibrium before  sphaleron decoupling,  which can increase  the lepton flavour asymmetry via the Yukawa couplings $h^*_{\alpha 2}h_{\beta 2}$ in~\eqref{eq:Salpha}.
   
For sterile neutrino masses below 1~GeV,  it is  more difficult to get viable points reaching the correct order of $Y_B$, as the corresponding $h$ couplings dictated by the seesaw relation become smaller. For masses above 50~GeV, washout effects and CP-violating sources from $L_{\rm tot}$-violating processes should also be included. Also, for heavier sterile neutrinos, lepto\-genesis through heavy neutrino decay to Higgs and leptons will  open. Without neutrino mass degeneracy, we expect lepton-flavour coherences to also boost CP-violating neutrino decay, contributing to freeze-out leptogenesis. A  general unifying framework that includes both light and heavy neutrinos will be developed elsewhere.

\medskip
\textbf{\textit{Conclusions}}. We have shown the importance of a dominant\- thermal resonant mechanism for low-scale lepto\-genesis due to thermally enhanced  lepton flavour coherences, without relying on sterile-neutrino mass degeneracies. The delineated sterile-neutrino mass spectrum for viable low-scale TIRL is different from the resonant and sterile-neutrino oscillation scenarios, and the general parameter space of TIRL in terms of the active-sterile mixing can be probed in laboratory experiments. 

\medskip
\textbf{\textit{Acknowledgements}}.	S.-P.~Li was supported by the JSPS Grant-in-Aid for JSPS Research Fellows No. 24KF0060, whilst AP's work is supported in part by
the STFC research grant: ST/X00077X/1.

\vspace*{-3mm}    
\bibliographystyle{JHEP}
\bibliography{Refs}

@misc{SM,
	note = {See more details in the Supplemental Material}
}

@article{Li:2026qym,
	author = "Li, Shao-Ping and Pilaftsis, Apostolos",
	title = "{Low-Scale Leptogenesis from Resonant Thermal Lepton Flavour Coherences}",
	eprint = "2604.06493",
	archivePrefix = "arXiv",
	primaryClass = "hep-ph",
	month = "4",
	year = "2026"
}

@article{CMS:2022fut,
    author = "Tumasyan, Armen and others",
    collaboration = "CMS",
    title = "{Search for long-lived heavy neutral leptons with displaced vertices in proton-proton collisions at $ \sqrt{\mathrm{s}} $ =13 TeV}",
    eprint = "2201.05578",
    archivePrefix = "arXiv",
    primaryClass = "hep-ex",
    reportNumber = "CMS-EXO-20-009, CERN-EP-2021-264",
    doi = "10.1007/JHEP07(2022)081",
    journal = "JHEP",
    volume = "07",
    pages = "081",
    year = "2022"
}

@article{ATLAS:2022atq,
    author = "Aad, Georges and others",
    collaboration = "ATLAS",
    title = "{Search for Heavy Neutral Leptons in Decays of W Bosons Using a Dilepton Displaced Vertex in s=13{\,}{\,}TeV pp Collisions with the ATLAS Detector}",
    eprint = "2204.11988",
    archivePrefix = "arXiv",
    primaryClass = "hep-ex",
    reportNumber = "CERN-EP-2022-017",
    doi = "10.1103/PhysRevLett.131.061803",
    journal = "Phys. Rev. Lett.",
    volume = "131",
    number = "6",
    pages = "061803",
    year = "2023"
}

@article{Dinh:2016q,
	author = "Dinh, Laurent and Sohl-Dickstein, Jascha and Bengio, Samy",
	title = "{Density estimation using real NVP}",
    note="In International
Conference on Learning Representations (2017)",
	eprint = "1605.08803",
	archivePrefix = "arXiv",
	primaryClass = "cs.LG",
	month = "5",
	year = "2016"  
}

@article{Esteban:2024eli,
    author = "Esteban, Ivan and Gonzalez-Garcia, M. C. and Maltoni, Michele and Martinez-Soler, Ivan and Pinheiro, Jo{\~a}o Paulo and Schwetz, Thomas",
    title = "{NuFit-6.0: updated global analysis of three-flavor neutrino oscillations}",
    eprint = "2410.05380",
    archivePrefix = "arXiv",
    primaryClass = "hep-ph",
    reportNumber = "IFT-UAM/CSIC-24-140, YITP-SB-2024-24, IPPP/24/64, IPPP/24/64, IFT-UAM/CSIC-24-140, YITP-SB-2024-24",
    doi = "10.1007/JHEP12(2024)216",
    journal = "JHEP",
    volume = "12",
    pages = "216",
    year = "2024"
}

@article{Carrington:1991hz,
    author = "Carrington, M. E.",
    title = "{The Effective potential at finite temperature in the Standard Model}",
    reportNumber = "TPI-MINN-91-48-T-REV, TPI-MINN-91-48-T",
    doi = "10.1103/PhysRevD.45.2933",
    journal = "Phys. Rev. D",
    volume = "45",
    pages = "2933--2944",
    year = "1992"
}

@article{Li:2025clq,
    author = "Li, Shao-Ping and Chluba, Jens",
    title = "{Neutrinogenic CMB spectral distortions}",
    eprint = "2510.04684",
    archivePrefix = "arXiv",
    primaryClass = "astro-ph.CO",
    reportNumber = "OU-HET 1292",
    month = "10",
    year = "2025"
}

@article{ParticleDataGroup:2024cfk,
    author = "Navas, S. and others",
    collaboration = "Particle Data Group",
    title = "{Review of particle physics}",
    doi = "10.1103/PhysRevD.110.030001",
    journal = "Phys. Rev. D",
    volume = "110",
    number = "3",
    pages = "030001",
    year = "2024"
}

@article{Canetti:2014dka,
	author = {Canetti, Laurent and Drewes, Marco and Garbrecht, Bj\"orn},
	title = "{Probing leptogenesis with GeV-scale sterile neutrinos at LHCb and Belle II}",
	eprint = "1404.7114",
	archivePrefix = "arXiv",
	primaryClass = "hep-ph",
	reportNumber = "TUM-HEP-939-14",
	doi = "10.1103/PhysRevD.90.125005",
	journal = "Phys. Rev. D",
	volume = "90",
	number = "12",
	pages = "125005",
	year = "2014"
}

@article{Canetti:2012kh,
    author = "Canetti, Laurent and Drewes, Marco and Frossard, Tibor and Shaposhnikov, Mikhail",
    title = "{Dark Matter, Baryogenesis and Neutrino Oscillations from Right Handed Neutrinos}",
    eprint = "1208.4607",
    archivePrefix = "arXiv",
    primaryClass = "hep-ph",
    reportNumber = "TTK-12-05, TUM-HEP-852-12, CAS-KITPC-ITP-368",
    doi = "10.1103/PhysRevD.87.093006",
    journal = "Phys. Rev. D",
    volume = "87",
    pages = "093006",
    year = "2013"
}

@article{Shuve:2014zua,
    author = "Shuve, Brian and Yavin, Itay",
    title = "{Baryogenesis through Neutrino Oscillations: A Unified Perspective}",
    eprint = "1401.2459",
    archivePrefix = "arXiv",
    primaryClass = "hep-ph",
    doi = "10.1103/PhysRevD.89.075014",
    journal = "Phys. Rev. D",
    volume = "89",
    number = "7",
    pages = "075014",
    year = "2014"
}

@article{Sigl:1993ctk,
    author = "Sigl, G. and Raffelt, G.",
    title = "{General kinetic description of relativistic mixed neutrinos}",
    reportNumber = "MPI-PH-92-112",
    doi = "10.1016/0550-3213(93)90175-O",
    journal = "Nucl. Phys. B",
    volume = "406",
    pages = "423--451",
    year = "1993"
}

@article{Chluba:2016bvg,
    author = "Chluba, Jens",
    title = "{Which spectral distortions does $\Lambda$CDM actually predict?}",
    eprint = "1603.02496",
    archivePrefix = "arXiv",
    primaryClass = "astro-ph.CO",
    doi = "10.1093/mnras/stw945",
    journal = "Mon. Not. Roy. Astron. Soc.",
    volume = "460",
    number = "1",
    pages = "227--239",
    year = "2016"
}

@article{Pilaftsis:2004xx,
    author = "Pilaftsis, Apostolos",
    title = "{Resonant tau-leptogenesis with observable lepton number violation}",
    eprint = "hep-ph/0408103",
    archivePrefix = "arXiv",
    doi = "10.1103/PhysRevLett.95.081602",
    journal = "Phys. Rev. Lett.",
    volume = "95",
    pages = "081602",
    year = "2005"
}

@article{BhupalDev:2014pfm,
    author = "Bhupal Dev, P. S. and Millington, Peter and Pilaftsis, Apostolos and Teresi, Daniele",
    title = "{Flavour Covariant Transport Equations: an Application to Resonant Leptogenesis}",
    eprint = "1404.1003",
    archivePrefix = "arXiv",
    primaryClass = "hep-ph",
    reportNumber = "MAN-HEP-2014-01, IPPP-14-20, DCPT-14-40",
    doi = "10.1016/j.nuclphysb.2014.06.020",
    journal = "Nucl. Phys. B",
    volume = "886",
    pages = "569--664",
    year = "2014"
}

@article{BhupalDev:2014oar,
    author = "Bhupal Dev, P. S. and Millington, Peter and Pilaftsis, Apostolos and Teresi, Daniele",
    title = "{Kadanoff{\textendash}Baym approach to flavour mixing and oscillations in resonant leptogenesis}",
    eprint = "1410.6434",
    archivePrefix = "arXiv",
    primaryClass = "hep-ph",
    reportNumber = "MAN-HEP-2014-13, TUM-HEP-962-14",
    doi = "10.1016/j.nuclphysb.2014.12.003",
    journal = "Nucl. Phys. B",
    volume = "891",
    pages = "128--158",
    year = "2015"
}

@article{Beneke:2010dz,
    author = "Beneke, Martin and Garbrecht, Bjorn and Fidler, Christian and Herranen, Matti and Schwaller, Pedro",
    title = "{Flavoured Leptogenesis in the CTP Formalism}",
    eprint = "1007.4783",
    archivePrefix = "arXiv",
    primaryClass = "hep-ph",
    reportNumber = "TTK-10-44, ZU-TH-09-10",
    doi = "10.1016/j.nuclphysb.2010.10.001",
    journal = "Nucl. Phys. B",
    volume = "843",
    pages = "177--212",
    year = "2011"
}

@article{Kersten:2007vk,
    author = {Kersten, J{\"o}rn and Smirnov, Alexei Yu.},
    title = "{Right-Handed Neutrinos at CERN LHC and the Mechanism of Neutrino Mass Generation}",
    eprint = "0705.3221",
    archivePrefix = "arXiv",
    primaryClass = "hep-ph",
    doi = "10.1103/PhysRevD.76.073005",
    journal = "Phys. Rev. D",
    volume = "76",
    pages = "073005",
    year = "2007"
}

@article{Boyarsky:2020dzc,
    author = "Boyarsky, Alexey and Ovchynnikov, Maksym and Ruchayskiy, Oleg and Syvolap, Vsevolod",
    title = "{Improved big bang nucleosynthesis constraints on heavy neutral leptons}",
    eprint = "2008.00749",
    archivePrefix = "arXiv",
    primaryClass = "hep-ph",
    doi = "10.1103/PhysRevD.104.023517",
    journal = "Phys. Rev. D",
    volume = "104",
    number = "2",
    pages = "023517",
    year = "2021"
}

@article{Datta:1993nm,
    author = "Datta, A. and Guchait, M. and Pilaftsis, A.",
    title = "{Probing lepton number violation via majorana neutrinos at hadron supercolliders}",
    eprint = "hep-ph/9311257",
    archivePrefix = "arXiv",
    reportNumber = "RAL-93-074",
    doi = "10.1103/PhysRevD.50.3195",
    journal = "Phys. Rev. D",
    volume = "50",
    pages = "3195--3203",
    year = "1994"
}

@article{Atre:2009rg,
    author = "Atre, Anupama and Han, Tao and Pascoli, Silvia and Zhang, Bin",
    title = "{The Search for Heavy Majorana Neutrinos}",
    eprint = "0901.3589",
    archivePrefix = "arXiv",
    primaryClass = "hep-ph",
    reportNumber = "FERMILAB-PUB-08-086-T, NSF-KITP-08-54, MADPH-06-1466, DCPT-07-198, IPPP-07-99",
    doi = "10.1088/1126-6708/2009/05/030",
    journal = "JHEP",
    volume = "05",
    pages = "030",
    year = "2009"
}

@article{Dev:2013wba,
    author = "Dev, P. S. Bhupal and Pilaftsis, Apostolos and Yang, Un-ki",
    title = "{New Production Mechanism for Heavy Neutrinos at the LHC}",
    eprint = "1308.2209",
    archivePrefix = "arXiv",
    primaryClass = "hep-ph",
    reportNumber = "MAN-HEP-2013-15, CERN-PH-TH-2013-192",
    doi = "10.1103/PhysRevLett.112.081801",
    journal = "Phys. Rev. Lett.",
    volume = "112",
    number = "8",
    pages = "081801",
    year = "2014"
}

@article{Deppisch:2015qwa,
    author = "Deppisch, Frank F. and Bhupal Dev, P. S. and Pilaftsis, Apostolos",
    title = "{Neutrinos and Collider Physics}",
    eprint = "1502.06541",
    archivePrefix = "arXiv",
    primaryClass = "hep-ph",
    reportNumber = "MAN-HEP-2014-15",
    doi = "10.1088/1367-2630/17/7/075019",
    journal = "New J. Phys.",
    volume = "17",
    number = "7",
    pages = "075019",
    year = "2015"
}

@article{Pilaftsis:1991ug,
    author = "Pilaftsis, Apostolos",
    title = "{Radiatively induced neutrino masses and large Higgs neutrino couplings in the standard model with Majorana fields}",
    eprint = "hep-ph/9901206",
    archivePrefix = "arXiv",
    reportNumber = "MZ-TH-91-32",
    doi = "10.1007/BF01482590",
    journal = "Z. Phys. C",
    volume = "55",
    pages = "275--282",
    year = "1992"
}

@article{Abdullahi:2022jlv,
    author = "Abdullahi, Asli M. and others",
    title = "{The present and future status of heavy neutral leptons}",
    eprint = "2203.08039",
    archivePrefix = "arXiv",
    primaryClass = "hep-ph",
    reportNumber = "FERMILAB-CONF-22-184-T-V",
    doi = "10.1088/1361-6471/ac98f9",
    journal = "J. Phys. G",
    volume = "50",
    number = "2",
    pages = "020501",
    year = "2023"
}

@article{FCC:2018evy,
    author = "Abada, A. and others",
    collaboration = "FCC",
    title = "{FCC-ee: The Lepton Collider}: {Future Circular Collider Conceptual Design Report Volume 2}",
    reportNumber = "CERN-ACC-2018-0057",
    doi = "10.1140/epjst/e2019-900045-4",
    journal = "Eur. Phys. J. ST",
    volume = "228",
    number = "2",
    pages = "261--623",
    year = "2019"
}

@article{CEPCStudyGroup:2018ghi,
    author = "Dong, Mingyi and others",
    editor = "Guimar{\~a}es da Costa, Jo{\~a}o Barreiro and others",
    collaboration = "CEPC Study Group",
    title = "{CEPC Conceptual Design Report: Volume 2 - Physics {\&} Detector}",
    eprint = "1811.10545",
    archivePrefix = "arXiv",
    primaryClass = "hep-ex",
    reportNumber = "IHEP-CEPC-DR-2018-02, IHEP-EP-2018-01, IHEP-TH-2018-01",
    month = "11",
    year = "2018"
}

@article{Antusch:2017pkq,
    author = "Antusch, Stefan and Cazzato, Eros and Drewes, Marco and Fischer, Oliver and Garbrecht, Bjorn and Gueter, Dario and Klaric, Juraj",
    title = "{Probing Leptogenesis at Future Colliders}",
    eprint = "1710.03744",
    archivePrefix = "arXiv",
    primaryClass = "hep-ph",
    reportNumber = "TUM-1160/18, CP3-17-48",
    doi = "10.1007/JHEP09(2018)124",
    journal = "JHEP",
    volume = "09",
    pages = "124",
    year = "2018"
}

@article{Antusch:2016ejd,
    author = "Antusch, Stefan and Cazzato, Eros and Fischer, Oliver",
    title = "{Sterile neutrino searches at future $e^-e^+$, $pp$, and $e^-p$ colliders}",
    eprint = "1612.02728",
    archivePrefix = "arXiv",
    primaryClass = "hep-ph",
    doi = "10.1142/S0217751X17500786",
    journal = "Int. J. Mod. Phys. A",
    volume = "32",
    number = "14",
    pages = "1750078",
    year = "2017"
}

@article{Drewes:2019fou,
    author = "Drewes, Marco and Hajer, Jan",
    title = "{Heavy Neutrinos in displaced vertex searches at the LHC and HL-LHC}",
    eprint = "1903.06100",
    archivePrefix = "arXiv",
    primaryClass = "hep-ph",
    reportNumber = "CP3-19-11",
    doi = "10.1007/JHEP02(2020)070",
    journal = "JHEP",
    volume = "02",
    pages = "070",
    year = "2020"
}

@article{Harz:2021psp,
    author = "Harz, Julia and Ramsey-Musolf, Michael J. and Shen, Tianyang and Quiroga, Sebasti{\'a}n Urrutia",
    title = "{TeV-scale lepton number violation: Connecting leptogenesis, neutrinoless double beta decay, and colliders}",
    eprint = "2106.10838",
    archivePrefix = "arXiv",
    primaryClass = "hep-ph",
    reportNumber = "ACFI-T21-08, TUM-HEP-1346/21",
    doi = "10.1103/PhysRevD.110.035024",
    journal = "Phys. Rev. D",
    volume = "110",
    number = "3",
    pages = "035024",
    year = "2024"
}

@article{Izaguirre:2015pga,
    author = "Izaguirre, Eder and Shuve, Brian",
    title = "{Multilepton and Lepton Jet Probes of Sub-Weak-Scale Right-Handed Neutrinos}",
    eprint = "1504.02470",
    archivePrefix = "arXiv",
    primaryClass = "hep-ph",
    doi = "10.1103/PhysRevD.91.093010",
    journal = "Phys. Rev. D",
    volume = "91",
    number = "9",
    pages = "093010",
    year = "2015"
}

@article{Mohapatra:1979ia,
    author = "Mohapatra, Rabindra N. and Senjanovic, Goran",
    title = "{Neutrino Mass and Spontaneous Parity Nonconservation}",
    reportNumber = "MDDP-TR-80-060, MDDP-PP-80-105, CCNY-HEP-79-10",
    doi = "10.1103/PhysRevLett.44.912",
    journal = "Phys. Rev. Lett.",
    volume = "44",
    pages = "912",
    year = "1980"
}

@article{Schechter:1980gr,
    author = "Schechter, J. and Valle, J. W. F.",
    title = "{Neutrino Masses in SU(2) x U(1) Theories}",
    reportNumber = "SU-4217-167, COO-3533-167",
    doi = "10.1103/PhysRevD.22.2227",
    journal = "Phys. Rev. D",
    volume = "22",
    pages = "2227",
    year = "1980"
}

@article{FASER:2018eoc,
    author = "Ariga, Akitaka and others",
    collaboration = "FASER",
    title = "{FASER{\textquoteright}s physics reach for long-lived particles}",
    eprint = "1811.12522",
    archivePrefix = "arXiv",
    primaryClass = "hep-ph",
    reportNumber = "UCI-TR-2018-19, KYUSHU-RCAPP-2018-06",
    doi = "10.1103/PhysRevD.99.095011",
    journal = "Phys. Rev. D",
    volume = "99",
    number = "9",
    pages = "095011",
    year = "2019"
}

@article{DUNE:2018tke,
    author = "Abi, B. and others",
    collaboration = "DUNE",
    title = "{The DUNE Far Detector Interim Design Report Volume 1: Physics, Technology and Strategies}",
    eprint = "1807.10334",
    archivePrefix = "arXiv",
    primaryClass = "physics.ins-det",
    reportNumber = "Fermilab-Design-2018-02, FERMILAB-DESIGN-2018-02",
    month = "7",
    year = "2018"
}

@article{NA62:2017rwk,
    author = "Cortina Gil, Eduardo and others",
    collaboration = "NA62",
    title = "{The Beam and detector of the NA62 experiment at CERN}",
    eprint = "1703.08501",
    archivePrefix = "arXiv",
    primaryClass = "physics.ins-det",
    doi = "10.1088/1748-0221/12/05/P05025",
    journal = "JINST",
    volume = "12",
    number = "05",
    pages = "P05025",
    year = "2017"
}

@article{Chou:2016lxi,
	author = "Chou, John Paul and Curtin, David and Lubatti, H. J.",
	title = "{New Detectors to Explore the Lifetime Frontier}",
	eprint = "1606.06298",
	archivePrefix = "arXiv",
	primaryClass = "hep-ph",
	doi = "10.1016/j.physletb.2017.01.043",
	journal = "Phys. Lett. B",
	volume = "767",
	pages = "29--36",
	year = "2017"
}

@article{Curtin:2018mvb,
	author = "Curtin, David and others",
	title = "{Long-Lived Particles at the Energy Frontier: The MATHUSLA Physics Case}",
	eprint = "1806.07396",
	archivePrefix = "arXiv",
	primaryClass = "hep-ph",
	reportNumber = "FERMILAB-PUB-18-264-T",
	doi = "10.1088/1361-6633/ab28d6",
	journal = "Rept. Prog. Phys.",
	volume = "82",
	number = "11",
	pages = "116201",
	year = "2019"
}

@article{SHiP:2018xqw,
	author = "Ahdida, C. and others",
	collaboration = "SHiP",
	title = "{Sensitivity of the SHiP experiment to Heavy Neutral Leptons}",
	eprint = "1811.00930",
	archivePrefix = "arXiv",
	primaryClass = "hep-ph",
	doi = "10.1007/JHEP04(2019)077",
	journal = "JHEP",
	volume = "04",
	pages = "077",
	year = "2019"
}

@article{Akhmedov:1998qx,
	author = "Akhmedov, Evgeny K. and Rubakov, V. A. and Smirnov, A. Yu.",
	title = "{Baryogenesis via neutrino oscillations}",
	eprint = "hep-ph/9803255",
	archivePrefix = "arXiv",
	reportNumber = "IC-98-22, INR-98-14-T",
	doi = "10.1103/PhysRevLett.81.1359",
	journal = "Phys. Rev. Lett.",
	volume = "81",
	pages = "1359--1362",
	year = "1998"
}

@article{Hambye:2016sby,
	author = "Hambye, Thomas and Teresi, Daniele",
	title = "{Higgs doublet decay as the origin of the baryon asymmetry}",
	eprint = "1606.00017",
	archivePrefix = "arXiv",
	primaryClass = "hep-ph",
	reportNumber = "ULB-TH-16-08",
	doi = "10.1103/PhysRevLett.117.091801",
	journal = "Phys. Rev. Lett.",
	volume = "117",
	number = "9",
	pages = "091801",
	year = "2016"
}

@article{Hambye:2017elz,
	author = "Hambye, Thomas and Teresi, Daniele",
	title = "{Baryogenesis from L-violating Higgs-doublet decay in the density-matrix formalism}",
	eprint = "1705.00016",
	archivePrefix = "arXiv",
	primaryClass = "hep-ph",
	reportNumber = "ULB-TH-17-07",
	doi = "10.1103/PhysRevD.96.015031",
	journal = "Phys. Rev. D",
	volume = "96",
	number = "1",
	pages = "015031",
	year = "2017"
}

@article{Asaka:2005an,
	author = "Asaka, Takehiko and Blanchet, Steve and Shaposhnikov, Mikhail",
	title = "{The nuMSM, dark matter and neutrino masses}",
	eprint = "hep-ph/0503065",
	archivePrefix = "arXiv",
	doi = "10.1016/j.physletb.2005.09.070",
	journal = "Phys. Lett. B",
	volume = "631",
	pages = "151--156",
	year = "2005"
}

@article{Abada:2018oly,
	author = "Abada, Asmaa and Arcadi, Giorgio and Domcke, Valerie and Drewes, Marco and Klaric, Juraj and Lucente, Michele",
	title = "{Low-scale leptogenesis with three heavy neutrinos}",
	eprint = "1810.12463",
	archivePrefix = "arXiv",
	primaryClass = "hep-ph",
	reportNumber = "CP3-18-59, DESY 18-174, DESY-18-174, LPT-Orsay-18-85",
	doi = "10.1007/JHEP01(2019)164",
	journal = "JHEP",
	volume = "01",
	pages = "164",
	year = "2019"
}

@article{Davidson:2008bu,
    author = "Davidson, Sacha and Nardi, Enrico and Nir, Yosef",
    title = "{Leptogenesis}",
    eprint = "0802.2962",
    archivePrefix = "arXiv",
    primaryClass = "hep-ph",
    doi = "10.1016/j.physrep.2008.06.002",
    journal = "Phys. Rept.",
    volume = "466",
    pages = "105--177",
    year = "2008"
}

@article{Cottin:2018nms,
    author = "Cottin, Giovanna and Helo, Juan Carlos and Hirsch, Martin",
    title = "{Displaced vertices as probes of sterile neutrino mixing at the LHC}",
    eprint = "1806.05191",
    archivePrefix = "arXiv",
    primaryClass = "hep-ph",
    reportNumber = "IFIC/18-25, IFIC-18-25",
    doi = "10.1103/PhysRevD.98.035012",
    journal = "Phys. Rev. D",
    volume = "98",
    number = "3",
    pages = "035012",
    year = "2018"
}

@article{Minkowski:1977sc,
	author = "Minkowski, Peter",
	title = "{$\mu \to e\gamma$ at a Rate of One Out of $10^{9}$ Muon Decays?}",
	reportNumber = "Print-77-0182 (BERN)",
	doi = "10.1016/0370-2693(77)90435-X",
	journal = "Phys. Lett. B",
	volume = "67",
	pages = "421--428",
	year = "1977"
}

@article{Yanagida:1980xy,
	author = "Yanagida, Tsutomu",
	title = "{Horizontal Symmetry and Masses of Neutrinos}",
	reportNumber = "TU-80-208",
	doi = "10.1143/PTP.64.1103",
	journal = "Prog. Theor. Phys.",
	volume = "64",
	pages = "1103",
	year = "1980"
}

@article{Fukugita:1986hr,
	author = "Fukugita, M. and Yanagida, T.",
	title = "{Baryogenesis Without Grand Unification}",
	reportNumber = "RIFP-641",
	doi = "10.1016/0370-2693(86)91126-3",
	journal = "Phys. Lett. B",
	volume = "174",
	pages = "45--47",
	year = "1986"
}

@article{Anisimov:2010gy,
	author = "Anisimov, Alexey and Besak, Denis and Bodeker, Dietrich",
	title = "{Thermal production of relativistic Majorana neutrinos: Strong enhancement by multiple soft scattering}",
	eprint = "1012.3784",
	archivePrefix = "arXiv",
	primaryClass = "hep-ph",
	reportNumber = "BI-TP-2010-48",
	doi = "10.1088/1475-7516/2011/03/042",
	journal = "JCAP",
	volume = "03",
	pages = "042",
	year = "2011"
}

@article{Laine:2013lka,
	author = "Laine, M.",
	title = "{Thermal right-handed neutrino production rate in the relativistic regime}",
	eprint = "1307.4909",
	archivePrefix = "arXiv",
	primaryClass = "hep-ph",
	doi = "10.1007/JHEP08(2013)138",
	journal = "JHEP",
	volume = "08",
	pages = "138",
	year = "2013"
}

@article{Garbrecht:2013bia,
	author = {Garbrecht, Bj\"orn and Glowna, Frank and Schwaller, Pedro},
	title = "{Scattering Rates For Leptogenesis: Damping of Lepton Flavour Coherence and Production of Singlet Neutrinos}",
	eprint = "1303.5498",
	archivePrefix = "arXiv",
	primaryClass = "hep-ph",
	reportNumber = "TUM-HEP-880-13, TTK-13-07, ANL-HEP-PR-13-19",
	doi = "10.1016/j.nuclphysb.2013.08.020",
	journal = "Nucl. Phys. B",
	volume = "877",
	pages = "1--35",
	year = "2013"
}

@article{Besak:2012qm,
	author = "Besak, Denis and Bodeker, Dietrich",
	title = "{Thermal production of ultrarelativistic right-handed neutrinos: Complete leading-order results}",
	eprint = "1202.1288",
	archivePrefix = "arXiv",
	primaryClass = "hep-ph",
	reportNumber = "BI-TP-2012-05",
	doi = "10.1088/1475-7516/2012/03/029",
	journal = "JCAP",
	volume = "03",
	pages = "029",
	year = "2012"
}

@article{Asaka:2005pn,
	author = "Asaka, Takehiko and Shaposhnikov, Mikhail",
	title = "{The $\nu$MSM, dark matter and baryon asymmetry of the universe}",
	eprint = "hep-ph/0505013",
	archivePrefix = "arXiv",
	doi = "10.1016/j.physletb.2005.06.020",
	journal = "Phys. Lett. B",
	volume = "620",
	pages = "17--26",
	year = "2005"
}

@article{Shi:1998km,
	author = "Shi, Xiang-Dong and Fuller, George M.",
	title = "{A New dark matter candidate: Nonthermal sterile neutrinos}",
	eprint = "astro-ph/9810076",
	archivePrefix = "arXiv",
	doi = "10.1103/PhysRevLett.82.2832",
	journal = "Phys. Rev. Lett.",
	volume = "82",
	pages = "2832--2835",
	year = "1999"
}

@article{Dodelson:1993je,
	author = "Dodelson, Scott and Widrow, Lawrence M.",
	title = "{Sterile-neutrinos as dark matter}",
	eprint = "hep-ph/9303287",
	archivePrefix = "arXiv",
	reportNumber = "FERMILAB-PUB-93-057-A",
	doi = "10.1103/PhysRevLett.72.17",
	journal = "Phys. Rev. Lett.",
	volume = "72",
	pages = "17--20",
	year = "1994"
}

@article{Bezrukov:2025ttd,
    author = "Bezrukov, Fedor and Gorbunov, Dmitry and Koreshkova, Ekaterina",
    title = "{Refining lower bounds on sterile neutrino dark matter mass from estimates of phase space densities in dwarf galaxies}",
    eprint = "2412.20585",
    archivePrefix = "arXiv",
    primaryClass = "hep-ph",
    doi = "10.1142/s0217751x25400044",
    journal = "Int. J. Mod. Phys. A",
    volume = "40",
    number = "33",
    pages = "2540004",
    year = "2025"
}

@article{Pilaftsis:1997jf,
	author = "Pilaftsis, Apostolos",
	title = "{CP violation and baryogenesis due to heavy Majorana neutrinos}",
	eprint = "hep-ph/9707235",
	archivePrefix = "arXiv",
	reportNumber = "MPI-PHT-97-30",
	doi = "10.1103/PhysRevD.56.5431",
	journal = "Phys. Rev. D",
	volume = "56",
	pages = "5431--5451",
	year = "1997"
}

@article{Pilaftsis:2003gt,
	author = "Pilaftsis, Apostolos and Underwood, Thomas E. J.",
	title = "{Resonant leptogenesis}",
	eprint = "hep-ph/0309342",
	archivePrefix = "arXiv",
	reportNumber = "MC-TH-2003-09",
	doi = "10.1016/j.nuclphysb.2004.05.029",
	journal = "Nucl. Phys. B",
	volume = "692",
	pages = "303--345",
	year = "2004"
}

@article{Pilaftsis:2005rv,
	author = "Pilaftsis, Apostolos and Underwood, Thomas E. J.",
	title = "{Electroweak-scale resonant leptogenesis}",
	eprint = "hep-ph/0506107",
	archivePrefix = "arXiv",
	doi = "10.1103/PhysRevD.72.113001",
	journal = "Phys. Rev. D",
	volume = "72",
	pages = "113001",
	year = "2005"
}

@article{Klaric:2020phc,
	author = "Klari\'c, Juraj and Shaposhnikov, Mikhail and Timiryasov, Inar",
	title = "{Uniting Low-Scale Leptogenesis Mechanisms}",
	eprint = "2008.13771",
	archivePrefix = "arXiv",
	primaryClass = "hep-ph",
	doi = "10.1103/PhysRevLett.127.111802",
	journal = "Phys. Rev. Lett.",
	volume = "127",
	number = "11",
	pages = "111802",
	year = "2021"
}

@article{Klaric:2021cpi,
	author = "Klari\'c, Juraj and Shaposhnikov, Mikhail and Timiryasov, Inar",
	title = "{Reconciling resonant leptogenesis and baryogenesis via neutrino oscillations}",
	eprint = "2103.16545",
	archivePrefix = "arXiv",
	primaryClass = "hep-ph",
	doi = "10.1103/PhysRevD.104.055010",
	journal = "Phys. Rev. D",
	volume = "104",
	number = "5",
	pages = "055010",
	year = "2021"
}

@article{Alekhin:2015byh,
	author = "Alekhin, Sergey and others",
	title = "{A facility to Search for Hidden Particles at the CERN SPS: the SHiP physics case}",
	eprint = "1504.04855",
	archivePrefix = "arXiv",
	primaryClass = "hep-ph",
	reportNumber = "CERN-SPSC-2015-017, SPSC-P-350-ADD-1",
	doi = "10.1088/0034-4885/79/12/124201",
	journal = "Rept. Prog. Phys.",
	volume = "79",
	number = "12",
	pages = "124201",
	year = "2016"
}

@article{Blondel:2014bra,
	author = "Blondel, Alain and Graverini, E. and Serra, N. and Shaposhnikov, M.",
	editor = "Aguilar-Ben{\'\i}tez, M and Fuster, J and Mart{\'\i}-Garc{\'\i}a, S and Santamar{\'\i}a, A",
	collaboration = "FCC-ee study Team",
	title = "{Search for Heavy Right Handed Neutrinos at the FCC-ee}",
	eprint = "1411.5230",
	archivePrefix = "arXiv",
	primaryClass = "hep-ex",
	doi = "10.1016/j.nuclphysbps.2015.09.304",
	journal = "Nucl. Part. Phys. Proc.",
	volume = "273-275",
	pages = "1883--1890",
	year = "2016"
}

@article{Drewes:2021nqr,
	author = "Drewes, Marco and Georis, Yannis and Klari\'c, Juraj",
	title = "{Mapping the Viable Parameter Space for Testable Leptogenesis}",
	eprint = "2106.16226",
	archivePrefix = "arXiv",
	primaryClass = "hep-ph",
	reportNumber = "CP3-21-43",
	doi = "10.1103/PhysRevLett.128.051801",
	journal = "Phys. Rev. Lett.",
	volume = "128",
	number = "5",
	pages = "051801",
	year = "2022"
}

@article{Giudice:2003jh,
	author = "Giudice, G. F. and Notari, A. and Raidal, M. and Riotto, A. and Strumia, A.",
	title = "{Towards a complete theory of thermal leptogenesis in the SM and MSSM}",
	eprint = "hep-ph/0310123",
	archivePrefix = "arXiv",
	reportNumber = "IFUP-TH-2003-37, CERN-TH-2003-240",
	doi = "10.1016/j.nuclphysb.2004.02.019",
	journal = "Nucl. Phys. B",
	volume = "685",
	pages = "89--149",
	year = "2004"
}

@article{DOnofrio:2014rug,
	author = "D'Onofrio, Michela and Rummukainen, Kari and Tranberg, Anders",
	title = "{Sphaleron Rate in the Minimal Standard Model}",
	eprint = "1404.3565",
	archivePrefix = "arXiv",
	primaryClass = "hep-ph",
	doi = "10.1103/PhysRevLett.113.141602",
	journal = "Phys. Rev. Lett.",
	volume = "113",
	number = "14",
	pages = "141602",
	year = "2014"
}

@article{Prokopec:2003pj,
	author = "Prokopec, Tomislav and Schmidt, Michael G. and Weinstock, Steffen",
	title = "{Transport equations for chiral fermions to order h bar and electroweak baryogenesis. Part 1}",
	eprint = "hep-ph/0312110",
	archivePrefix = "arXiv",
	reportNumber = "BNL-72343-2004-JA, HD-THEP-03-62",
	doi = "10.1016/j.aop.2004.06.002",
	journal = "Annals Phys.",
	volume = "314",
	pages = "208--265",
	year = "2004"
}

@article{Prokopec:2004ic,
	author = "Prokopec, Tomislav and Schmidt, Michael G. and Weinstock, Steffen",
	title = "{Transport equations for chiral fermions to order h-bar and electroweak baryogenesis. Part II}",
	eprint = "hep-ph/0406140",
	archivePrefix = "arXiv",
	reportNumber = "BNL-72342-2004-JA, HD-THEP-04-22",
	doi = "10.1016/j.aop.2004.06.001",
	journal = "Annals Phys.",
	volume = "314",
	pages = "267--320",
	year = "2004"
}

@article{Casas:2001sr,
	author = "Casas, J. A. and Ibarra, A.",
	title = "{Oscillating neutrinos and $\mu \to e, \gamma$}",
	eprint = "hep-ph/0103065",
	archivePrefix = "arXiv",
	reportNumber = "IEM-FT-211-01, OUTP-01-11P, IFT-UAM-CSIC-01-08",
	doi = "10.1016/S0550-3213(01)00475-8",
	journal = "Nucl. Phys. B",
	volume = "618",
	pages = "171--204",
	year = "2001"
}

@article{Harvey:1990qw,
	author = "Harvey, Jeffrey A. and Turner, Michael S.",
	title = "{Cosmological baryon and lepton number in the presence of electroweak fermion number violation}",
	reportNumber = "FERMILAB-PUB-90-049-A, EFI-90-33",
	doi = "10.1103/PhysRevD.42.3344",
	journal = "Phys. Rev. D",
	volume = "42",
	pages = "3344--3349",
	year = "1990"
}

@article{Planck:2018vyg,
	author = "Aghanim, N. and others",
	collaboration = "Planck",
	title = "{Planck 2018 results. VI. Cosmological parameters}",
	eprint = "1807.06209",
	archivePrefix = "arXiv",
	primaryClass = "astro-ph.CO",
	doi = "10.1051/0004-6361/201833910",
	journal = "Astron. Astrophys.",
	volume = "641",
	pages = "A6",
	year = "2020",
	note = "[Erratum: Astron.Astrophys. 652, C4 (2021)]"
}

\onecolumngrid

\pagebreak

\setcounter{page}{1}

	\begin{center}
	{\Large\textbf{Supplemental Material}}
\end{center}
\begin{center}
	{\large\textit{Dominant Thermal Resonant Mechanism for Low-Scale Leptogenesis}}
\end{center}
\begin{center}
	{\large  Shao-Ping Li and Apostolos Pilaftsis}
\end{center}

\section{Thermally Complete Transport Equations in the Density-Matrix Formalism}\label{sec:full-kin-eq}

In this appendix, we present a thermally complete set of heavy-neutrino and lepton number transport equations in the leading approximation of the density-matrix formalism for all $1 \leftrightarrow 2$ rates. Specifically, we consider all decays and inverse decays that are kinematically allowed in the on-shell (OS) quasi-thermal approximation. We expect that this simplified effective framework will capture the leading phenomena of quantum flavour (de)-coherences in both the sterile neutrino and active lepton sectors.

In line with the flavour covariant formalism in~\cite{BhupalDev:2014pfm}, we analogously define the density-number matrices for particle (anti-particle) species:
\begin{equation}
	\label{eq:nNLE}
	n^N\, (\bar{n}^N)\: \equiv\: \{ n^N_{ij}\, (\bar{n}^N_{ij})\}\,,\qquad  
	n^L\, (\bar{n}^L)\: \equiv\: \{n^L_{\!\alpha\beta}\, (\bar{n}^L_{\alpha\beta}) \}\,,\qquad n^E\, (\bar{n}^E)\: \equiv\: \{ n^E_{\!\alpha'\beta'}\, (\bar{n}^E_{\alpha'\beta'})\}\,,    
\end{equation}
which describe $N_N$ heavy neutrinos $N_{\!i,j}$ (with $i,j=1,2,\dots N_N$), $N_L$ left-handed lepton doublets $L_{\alpha,\beta}$ (with $\alpha,\beta =1,2,\dots N_L$), and $N_E = N_L$ right-handed leptons $E_{\alpha',\beta'}$ (with $\alpha',\beta'=1,2,\dots N_E$), respectively. Evidently, the complete leptonic sector exhibits flavour covariance under the product group
\begin{equation}
	\label{eq:Gflavour}
	G_\text{F}\: =\:  U(N_N)_N\times U(N_L)_L \times U(N_E)_E\,,   
\end{equation}
where ${N_L = N_E}$ for the minimal extension of the SM under study. As far as the number densities of the SM scalar doublet $\Phi$, and the electroweak gauge bosons $W^{1,2,3}_\mu$ and $B^\mu$ are concerned, we assume that these are given by their thermal equilibrium values to a good approximation. 

In addition, we define the thermally averaged self-energy matrices for all fermionic $X \in \{N_i\,, L_\alpha\,, E_{\alpha'}\}$ species:
\begin{equation}
	\label{eq:Eself}
	{\cal E}_X \: \equiv\: g_X\, \big(n^X_{\rm eq}\big)^{-1} \int_{\bf p}\, E_X({\bf p})\, \Big( e^{E_X ({\bf p})/T} + {\bf 1}\Big)^{-1}\,,
\end{equation}
where $E_X ({\bf p})$ is the self-energy matrix in the 
$X$-flavour space and the equilibrium number densities are given by
\begin{equation}
	\label{eq:nXeq} 
	n^X_{\rm eq}\: =\:  g_X \int_{\bf p}\,\Big( e^{E_X ({\bf p})/T} + {\bf 1}\Big)^{-1}\,,
\end{equation}
with $\int_{\bf p}\equiv \int {\rm d}^3{\bf p}/(2\pi)^3$, and $g_{N_i} = 2$, $g_{L_\alpha} = 4$ and $g_{E_\alpha'} = 2$  the relativistic degrees of freedom in the thermal plasma. Moreover, as done in~\cite{BhupalDev:2014pfm}, it proves convenient to define the generalised flavour-covariant collision terms for decays and inverse decays. In the OS quasi-thermal approximation, these have the following flavour and kinematic structure:
\begin{eqnarray}
	\label{eq:ColTerms}
	[\gamma^N_{L\Phi}]^{ij}_{\alpha\beta} &\equiv& [\gamma (N \to L\widetilde{\Phi}]^{ij}_{\alpha\beta}\: \propto\: {\bf h}^N_{\alpha i} {\bf h}^{N*}_{\beta j}\, \theta(m_{N_i} - \tilde{m}_{L_\alpha} - \tilde{M}_\Phi )\, \theta (m_{N_j} - \tilde{m}_{L_\beta} - \tilde{M}_\Phi )\, ,\\[0.2cm]
	{}[\gamma^{\bar{N}}_{\bar{L}\bar{\Phi}}]^{ij}_{\alpha\beta} &\equiv & [\gamma (\bar{N} \to \bar{L}\widetilde{\Phi}^\dagger) ]^{ij}_{\alpha\beta}\: \propto\: \bar{\bf h}^N_{\alpha i} \bar{\bf h}^{N*}_{\beta j}\, \theta(m_{N_i} - \tilde{m}_{L_\alpha} - \tilde{M}_\Phi )\, \theta (m_{N_j} - \tilde{m}_{L_\beta} - \tilde{M}_\Phi )\, ,\\[0.2cm]
	{}[\gamma^{\bar{\Phi}}_{\bar{N}L}]_{ij,\alpha\beta} &\equiv & [\gamma (\widetilde{\Phi}^\dagger \to \bar{N} L)]_{ij,\alpha\beta}\: \propto\: {\bf h}^{\Phi}_{\alpha i} {\bf h}^{\Phi *}_{\beta j}\, \theta(\tilde{M}_\Phi - m_{N_i} - \tilde{m}_{L_\alpha})\, \theta (\tilde{M}_\Phi - m_{N_j} - \tilde{m}_{L_\beta} )\,,\\[0.2cm]
	{}[\gamma^\Phi_{N\bar{L}}]_{ij,\alpha\beta} &\equiv & [\gamma (\widetilde{\Phi} \to N\bar{L})]_{ij,\alpha\beta}\: \propto\: \bar{\bf h}^{\Phi}_{\alpha i} \bar{\bf h}^{\Phi *}_{\beta j}\, \theta(\tilde{M}_\Phi - m_{N_i} - \tilde{m}_{L_\alpha})\, \theta (\tilde{M}_\Phi - m_{N_j} - \tilde{m}_{L_\beta} )\,,\\[0.2cm]
	{}[\gamma^{\Phi}_{\bar{E}L}]_{\alpha'\beta',\alpha\beta} &\equiv & [\gamma (\widetilde{\Phi} \to \bar{E}L)]_{\alpha'\beta',\alpha\beta}\: \propto\: {\bf y}_{\alpha \alpha'} {\bf y}^*_{\beta \beta'}\, \theta(\tilde{M}_\Phi - \tilde{m}_{E_{\alpha'}}\! - \tilde{m}_{L_\alpha})\, \theta (\tilde{M}_\Phi - \tilde{m}_{E_{\beta'}}\! - \tilde{m}_{L_\beta} )\,,\\[0.2cm]
	{}[\gamma^{\bar{\Phi}}_{E\bar{L}}]_{\alpha'\beta',\alpha\beta} &\equiv & [\gamma (\widetilde{\Phi}^\dagger \to E\bar{L})]_{\alpha'\beta',\alpha\beta}\: \propto\: \bar{\bf y}_{\alpha \alpha'} \bar{\bf y}^*_{\beta \beta'}\, \theta(\tilde{M}_\Phi - \tilde{m}_{E_{\alpha'}}\! - \tilde{m}_{L_\alpha})\, \theta (\tilde{M}_\Phi - \tilde{m}_{E_{\beta'}}\! - \tilde{m}_{L_\beta} )\,,\\[0.2cm]
	\label {eq:gamWmu}
	{}[\gamma^{W_\mu}_{L\bar{L}}]_{\beta \delta,\alpha\gamma} &\equiv & [\gamma (W_\mu\to L\bar{L})]_{\beta \delta,\alpha\gamma}\: \propto\: g_{W^I}g_{W^J}\, {\bf g}^W_{\alpha \beta}\, {\bf g}^{W*}_{\gamma \delta}\, \theta(\tilde{M}_{W^I} - \tilde{m}_{L_\alpha} - \tilde{m}_{L_\beta} )\, \theta (\tilde{M}_{W^J} - \tilde{m}_{L_\gamma}
	- \tilde{m}_{L_\delta}  )\quad \text{etc.}\quad
\end{eqnarray}
In the above expressions, $\theta (z)$ is the usual Heaviside step function, and the quantities ${\bf h}^{N,\Phi} = {\bf h}^{N,\Phi}(h,iA)$ are resummed effective couplings dominated by the absorptive parts of the self-energy  (potentially thermal) corrections that are generically denoted as $iA$. At tree-level, we have ${\bf h}^N_{\alpha i} = {\bf h}^\Phi_{\alpha i} = h_{\alpha i}$. The bar over these quantities indicate CP conjugation, which is obtained 
from ${\bf h}^{N,\Phi}$ as $\bar{\bf h}^{N,\Phi} = {\bf h}^{N,\Phi}(h^*,iA^*) \neq ({\bf h}^{N,\Phi})^*$, so $\bar{\bf h}^{N} = \bar{\bf h}^{\Phi} = h^*$ holds only at tree level. In a similar fashion, we can define
effective Yukawa couplings for the charged leptons ${\bf y}_{\alpha\alpha'}$ and gauge bosons ${\bf g}^W_{\alpha\beta}$. 
In the Born approximation, it is ${\bf g}^W_{\alpha\beta}=\delta_{\alpha\beta}$ and ${\bf y}_{\alpha\alpha'} = y_\alpha \delta_{\alpha\alpha'}$ in the weak basis in which the charged-lepton Yukawa matrix is diagonal. 
Furthermore, the symbol tilde $(\,\tilde{}\,)$ indicates that the masses are thermally corrected from SM interactions only. Additionally, we  take $m_{N_i} \simeq M_i$ to a good approximation, where $M_i$ are the singlet neutrino masses after electroweak symmetry breaking. 
Finally, in~\eqref{eq:gamWmu}, $W^I_\mu = W^{0,1,2,3}_\mu$ (with $W^0_\mu \equiv B_\mu)$  collectively represents the electroweak gauge bosons of U(1)$_Y$ and SU(2)$_{\rm L}$, whose gauge couplings are: $g_{W^0} \equiv g'$ and $g_{W^{1,2,3}} \equiv g_w$.

Due to the CPT symmetry of the theory, the following relations among the collision terms may be established:
\begin{equation}
	\label{eq:CPTrel}
	\gamma^N_{L\Phi}\: =\: \gamma_{\bar{N}}^{\bar{L}\bar{\Phi}}\,,\quad 
	\gamma^{\bar{N}}_{\bar{L}\bar{\Phi}}\: =\: \gamma_{N}^{L\Phi}\,,\quad 
	\gamma^\Phi_{N\bar{L}}\: =\: \gamma_{\bar{\Phi}}^{\bar{N}L}\,,\quad 
	\gamma^{\bar{\Phi}}_{E\bar{L}}\: =\: \gamma_{\Phi}^{\bar{E}L}\,,
	\quad 
	\gamma^{W_\mu}_{L\bar{L}}\: =\: \gamma_{W_\mu}^{L\bar{L}}\quad {\rm etc.}
\end{equation}
Note that our formalism includes resonant mixing and resonant coherent transitions between heavy-neutrino and charged-lepton states, which become more pronounced for small mass differences of heavy neutrinos
and charged leptons in the $\mu$-$e$ sector. The latter effect, namely the thermally induced transitions or coherences between the left-handed lepton doublets, is the central novel aspect presented in this Letter. As we shall see in the numerical analysis, this effect can dramatically change the predictions for BAU from the usual ARS mechanism for a wide range of heavy neutrino masses and their couplings.

Equipped with the above definitions and short-hand notations, the transport equations for the $N$-$L$-$E$ system may be written down in a concise matrix form as follows:
\begin{eqnarray}
	\label{eq:dotnN}
	\dot{n}^N &=& -i\Big[\,{\cal E}_N\,, n^N\Big]\: -\: 
	\frac{1}{2n^N_{\rm eq}}\,
	\Big\{ n^N\,, \text{Tr}_L\big(\gamma^N_{L\Phi}\big)\Big\}\: +\: 
	\frac{1}{n^L_{\rm eq}}\, \text{Tr}_{L}\big(n^L\gamma^{L\Phi}_N\big) 
	-\: \frac{1}{2n^N_{\rm eq}n^L_{\rm eq}}\,
	\Big\{ n^N\,, \text{Tr}_{\bar{L}}\big(\bar{\gamma}^{\bar{N}L}_{\bar{\Phi}} \bar{n}^L\big)\Big\}\: +\:  \text{Tr}_{\bar{L}}\big(\bar{\gamma}^{\bar{\Phi}}_{\bar{N}L}\big)\,,\qquad\\[3mm]
	\label{eq:dotnbarN}
	\dot{\bar n}^N &=& +i\Big[\,{\cal E}_N\,, \bar{n}^N\Big]\: -\: 
	\frac{1}{2n^N_{\rm eq}}\,
	\Big\{ \bar{n}^N\,, \text{Tr}_{\bar{L}}
	\big(\bar{\gamma}^{N}_{L\Phi}\big)\Big\}\: +\: 
	\frac{1}{n^L_{\rm eq}}\, \text{Tr}_{\bar{L}}\big(\bar{n}^L\bar{\gamma}^{L\Phi}_{N}
	\big) 
	-\: \frac{1}{2n^N_{\rm eq}n^L_{\rm eq}}\,
	\Big\{ \bar{n}^N\,, \text{Tr}_{L}\big( \gamma^{\bar{N}L}_{\bar{\Phi}} n^L\big)\Big\}\: +\:  \text{Tr}_{L}\big(\gamma^{\bar{\Phi}}_{\bar{N}L}\big)\,,\\[3mm]
	\label{eq:dotnL}
	\dot{n}^L &=& -i\Big[\,{\cal E}_L\,, n^L\Big]\: -\: 
	\frac{1}{2n^L_{\rm eq}}\,
	\Big\{ n^L\,, \text{Tr}_N\big(\gamma_N^{L\Phi}\big)\Big\}\: +\: 
	\frac{1}{n^N_{\rm eq}}\, \text{Tr}_{N}\big(n^N\gamma_{L\Phi}^N\big) 
	-\: \frac{1}{2n^N_{\rm eq}n^L_{\rm eq}}\,
	\Big\{ n^L\,, \text{Tr}_{\bar{N}}\big(\bar{n}^N \gamma^{\bar{N}L}_\Phi \big)\Big\}\: +\:  \text{Tr}_{\bar{N}}\big(\gamma^{\bar{\Phi}}_{\bar{N}L}\big)\nonumber\\
	&&-\: \frac{1}{2n^L_{\rm eq}n^E_{\rm eq}}\,
	\Big\{ n^L\,, \text{Tr}_{\bar{E}}\big(\bar{n}^E \gamma^{\bar{E}L}_{\bar{\Phi}} \big)\Big\}\: +\:  \text{Tr}_{\bar{E}}\big(\gamma^{\bar{\Phi}}_{\bar{E}L}\big)
	-\: \frac{1}{2n^L_{\rm eq}}\,
	\Big\{ n^L\,, \text{Tr}_{\bar{L}}\big(\gamma^{L\bar{L}}_{W_\mu} \bar{n}^L \big)\Big\}\: +\:  \text{Tr}_{\bar{L}}\big(\gamma^{W_\mu}_{L\bar{L}}\big)\,,\\[3mm]
	\label{eq:dotnbarL}
	\dot{\bar n}^L &=& +i\Big[\,{\cal E}_L\,, \bar{n}^L\Big]\: -\: 
	\frac{1}{2n^L_{\rm eq}}\,
	\Big\{ \bar{n}^L\,, \text{Tr}_{\bar{N}}\big(\bar{\gamma}_{N}^{L\Phi}\big)\Big\}\: +\: 
	\frac{1}{n^N_{\rm eq}}\, \text{Tr}_{\bar{N}}\big(\bar{n}^N\bar{\gamma}_{L\Phi}^{N}\big) 
	-\: \frac{1}{2n^N_{\rm eq}n^L_{\rm eq}}\,
	\Big\{ \bar{n}^L\,, \text{Tr}_{N}\big(n^N \bar{\gamma}^{\bar{N}L}_{\Phi} \big)\Big\}\: +\:  \text{Tr}_{N}\big(\bar{\gamma}^{\bar{\Phi}}_{\bar{N}L}\big)\nonumber\\
	&&-\: \frac{1}{2n^L_{\rm eq}n^E_{\rm eq}}\,
	\Big\{ \bar{n}^L\,, \text{Tr}_{E}\big(n^E \bar{\gamma}^{\bar{E}L}_{\bar{\Phi}} \big)\Big\}\: +\:  \text{Tr}_{E}\big(\bar{\gamma}^{\bar{\Phi}}_{\bar{E}L}\big)
	-\: \frac{1}{2n^L_{\rm eq}}\,
	\Big\{ \bar{n}^L\,, \text{Tr}_L\big( n^L\bar{\gamma}^{L\bar{L}}_{W_\mu} \big)\Big\}\: +\:  \text{Tr}_{L}\big(\bar{\gamma}^{W_\mu}_{L\bar{L}}\big)\,,\\[3mm]
	\label{eq:dotnE}
	\dot{n}^E &=& -i\Big[\,{\cal E}_E\,, n^E\Big]\: 
	-\: \frac{1}{2n^E_{\rm eq}n^L_{\rm eq}}\,
	\Big\{ n^E\,, \text{Tr}_{\bar{L}}\big(
	\bar{\gamma}^{\bar{E}L}_{\bar{\Phi}}\bar{n}^L\big)\Big\}\: +\:  \text{Tr}_{\bar{L}}\big(\bar{\gamma}^{\bar{E}L}_{\bar{\Phi}}\big)\:
	-\: \frac{1}{2n^E_{\rm eq}}\,
	\Big\{ n^E\,, \text{Tr}_{\bar{E}}\big( \gamma^{E\bar{E}}_{B_\mu}\, \bar{n}^E\big)\Big\}\: +\:  
	\text{Tr}_{\bar{E}}\big(\gamma^{B_\mu}_{E\bar{E}}\big)\,,\\[3mm]
	\label{eq:dotnbarE}
	\dot{\bar n}^E &=& +i\Big[\,{\cal E}_E\,, \bar{n}^E\Big]\: -\: \frac{1}{2n^E_{\rm eq}n^L_{\rm eq}}\,
	\Big\{ \bar{n}^E\,, \text{Tr}_{L}\big(\gamma^{\bar{E}L}_{\bar{\Phi}}\,
	n^L\big)\Big\}\: +\:  \text{Tr}_{L}\big(\gamma^{\bar{E}L}_{\bar{\Phi}}\big)\:
	-\: \frac{1}{2n^E_{\rm eq}}\,
	\Big\{ \bar{n}^E\,, \text{Tr}_{E}\big( n^E\bar{\gamma}^{E\bar{E}}_{B_\mu} \big)\Big\}\: 
	+\:  \text{Tr}_{E}\big(\bar{\gamma}^{B_\mu}_{E\bar{E}}\big)\,.
\end{eqnarray}
Here, $\dot{n}^X \equiv {\rm d}n^X/{\rm d}t$ denotes differentiation with respect to cosmic time $t$, with $X=N,L,E$, and $\text{Tr}_X(\cdots )$ or $\text{Tr}_{\bar{X}}(\cdots )$ stands for taking the trace {\it only} over flavour indices that run in the flavour space $X$ or $\bar{X}$. In~\eqref{eq:dotnbarN}, \eqref{eq:dotnbarL} and~\eqref{eq:dotnbarE}, the newly introduced collision terms $\bar{\gamma}^{X}_{YZ}$ are defined as $\bar{\gamma}^{X}_{YZ} \equiv \big(\gamma^{\bar{X}}_{\bar{Y}\bar{Z}}\big)^{\sf T}$ (with $X,Y,Z \in \{N\,, L\,, E\,, \Phi\}$), where the transposition on the CP-conjugate collision term $\gamma^{\bar{X}}_{\bar{Y}\bar{Z}}$ acts on all involved fermionic $N$-, $L$-and $E$-flavour spaces~\cite{BhupalDev:2014pfm}.

Some remarks are now in order concerning our approximations and other subtleties which were taken into account when deriving the transport equations~\eqref{eq:dotnN}--\eqref{eq:dotnbarE}:
\begin{itemize}
	
	\item[(i)] In thermal equilibrium, the number densities  of the particle species $X = N,L,E$ and their antiparticle counterparts~$\bar{X}$ are equal with vanishing chemical potentials, 
	i.e.~$n^X_{\rm eq} = \bar{n}^X_{\rm eq}$ [cf.~\eqref{eq:nXeq}]. Hence, ${\cal E}_X$ and $n^X_{\rm eq}$ can be diagonalised in the same weak basis in which $E_X({\bf p}_X)$ is diagonal in the OS quasi-thermal approximation. Similarly, since all electroweak gauge bosons $W^I_\mu$ are in thermal equilibrium, we take their number densities to be diagonal: $(n^W_{\rm eq})_{IJ} \propto \delta_{IJ}$. 
	
	\item[(ii)] In the OS quasi-thermal approximation, we note that the collision terms $\gamma^N_{L\Phi}$ and $\gamma^{\bar{\Phi}}_{\bar{N}L}$ cannot occur simultaneously, since they are not thermally allowed at the same time.
	Moreover, to avoid an ambiguity due to the operation of transposition, all collision terms $\gamma^X_{YZ}$ or $\gamma_X^{YZ}$ have been uniformly defined with respect to $L$ in all six transport equations.    
	
	\item[(iii)] The commutator $[\,{\cal E}_X\,, n^X]$ (with $X=N,L,E$) is strictly valid for a single value of the three-momentum ${\bf p}_X$, before integrating over the entire phase space of the species $X$. 
	However, as was shown in detail in~\cite{Li:2026qym} within the Kadanoff-Baym formalism, $[\,{\cal E}_X\,, n^X]$ is still a good approximation for temperatures~$T\sim 200$~GeV that lie above the sphaleron freeze-out temperature $T_{\rm sph}\approx 132$~GeV in the~SM~\cite{DOnofrio:2014rug}, when $m_{N_i}\ll T_{\rm sph}$. Thus, this approximation offers a significant simplification in order to obtain credible numerical estimates for the BAU within scenarios of our interest here. We may remind the reader that in standard resonant leptogenesis (RL) with $m_{N_i}\stackrel{>}{{}_\sim} T_{\rm sph}$, the high degeneracy of heavy neutrinos ensures the validity of this approximation~\cite{BhupalDev:2014pfm}.
	
	\item[(iv)] Besides the Fermi--Dirac distribution for the fermionic $X$-sector particles and the Bose--Einstein distribution for the scalar doublet $\Phi$, we have not included for illustration other statistical factors in our transport equations, such as Pauli-blocking factors, for the final-state particles. However, their inclusion is straightforward and these additional factors have been implemented in~the numerical part of the present study and in~\cite{Li:2026qym} within the Kadanoff-Baym formalism.
	
	\item[(v)] Unlike previous studies of
	leptogenesis (e.g.~see~\cite{BhupalDev:2014pfm}), we do not assume {\it ab initio} that $n^L + \bar{n}^L = 2\,n^L_{\rm eq}$ in all our derivations of the transport equations in this appendix.

\end{itemize}

In order to assess the significance of quantum (de)-coherence from each $X$ sector and their contribution to the BAU, we should translate the transport equations~\eqref{eq:dotnN}--\eqref{eq:dotnbarE} in an eigenbasis of CP$\widetilde{\text{T}}$, where $\widetilde{\text{T}}$ denotes {\it naive} time-reversal transformation.  Under~$\widetilde{\text{T}}$, the cosmic time $t$ remains the same, but three-momenta and spins of particles change sign. Moreover, the number densities~$n^X$ get transposed under a~$\widetilde{\text{T}}$ transformation:  $n^X \to (n^X)^{\widetilde{\text{T}}} = (n^X)^{\sf T}$. More explicitly, under the action of CP and CP$\widetilde{\text{T}}$, we~have
\begin{equation}
	\label{eq:CPTtilde}
	\text{CP}:\ \  n^X\: \to\: (n^X)^{\rm CP} = (\bar{n}^X)^{\sf T}\,,\qquad
	{\rm CP}\widetilde{\text{T}}:\ \ n^X\: \to\: (n^X)^{\rm CP\widetilde{\rm T}} = \bar{n}^X\,.
\end{equation}
With the help of~\eqref{eq:CPTtilde}, we may now define the CP$\widetilde{\text{T}}$-even and CP$\widetilde{\text{T}}$-odd number density matrices
\begin{equation}
	\label{eq:numberCPT}
	\underline{n}^X\: \equiv\: \frac{1}{2}\, \Big( n^X + \bar{n}^X\Big)\,,\qquad 
	\Delta n^X\: \equiv\: n^X - \bar{n}^X\,,
\end{equation}
respectively. By analogy, we define the flavour-covariant 
CP$\widetilde{\text{T}}$-even and CP$\widetilde{\text{T}}$-odd collision terms
\begin{equation}
	\label{eq:gammaXYZ}
	\underline{\gamma}^X_{YZ}\: \equiv\: \frac{1}{2}\,\Big(\gamma^{X}_{YZ} + \bar{\gamma}^{X}_{YZ}\Big)\,,\qquad 
	\delta \gamma^X_{YZ}\: \equiv\: \gamma^{X}_{YZ} -\bar{\gamma}^{X}_{YZ}\,.
\end{equation}
Note that $\bar{\gamma}^{X}_{YZ} \equiv \big(\gamma^{\bar{X}}_{\bar{Y}\bar{Z}}\big)^{\sf T}$ is the proper CP$\widetilde{\text{T}}$-conjugate transform of $\gamma^{X}_{YZ}$, where the transposition on the CP-conjugate collision term $\gamma^{\bar{X}}_{\bar{Y}\bar{Z}}$
acts on both fermionic $N$- and $L$-flavour spaces~\cite{BhupalDev:2014pfm}, and similarly for the
$E$- and $L$-flavour spaces. As a consequence of their definitions, the CP$\widetilde{\text{T}}$-even and CP$\widetilde{\text{T}}$-odd  collision terms, $\underline{\gamma}^X_{YZ}$ and $\delta \gamma^X_{YZ}$, satisfy the following properties:
\begin{eqnarray}
	\label{eq:CPTtilde1}
	\underline{\gamma}^{\bar{X}}_{\bar{Y}\bar{Z}} \!&=&\! \big(\underline{\gamma}^X_{YZ}\big)^{\sf T}\;,\hspace{19.5mm}
	\delta\gamma^{\bar{X}}_{\bar{Y}\bar{Z}} \: =\: -\, \big(\delta\gamma^X_{YZ}\big)^{\sf T}\;,\\[0.2cm]
	\label{eq:CPTtilde2}
	\underline{\gamma}^{X}_{Y Z} \!&=&\! \underline{\gamma}_{\bar{X}}^{\bar{Y}\bar{Z}}\: =\: \big(\underline{\gamma}_{X}^{Y Z}\big)^{\sf T}\,,\qquad 
	\delta\gamma^{X}_{Y Z} \: =\: \delta\gamma_{\bar{X}}^{\bar{Y}\bar{Z}}\: =\: -\,\big(\delta\gamma_{X}^{Y Z}\big)^{\sf T}\,.
\end{eqnarray}
Observe that at the tree level, it is $\underline{\gamma}^{X}_{Y Z} = \gamma^{X}_{Y Z} = \bar{\gamma}^{X}_{Y Z}$, implying $\delta\gamma^{X}_{Y Z}=0$, e.g.~$\delta\gamma^{\bar{\Phi}}_{\bar{N}L} = 0$. At the loop level, one may have $\delta\gamma^{X}_{Y Z}\ne 0$ due to the absorptive parts of radiative corrections, as it happens in ordinary leptogenesis models~\cite{Fukugita:1986hr}, i.e.~${\delta\gamma^{N}_{L\Phi}\ne 0}$,
as well as in low-scale RL scenarios~\cite{Pilaftsis:2003gt}, with $\delta\gamma^{\bar{\Phi}}_{\bar{N}L} \ne 0$~\cite{Hambye:2016sby}. Hence, the present improved formulation allows a clear separation between radiative mixing and tree-level oscillation pheno\-mena among fermionic states. 

In the CP$\widetilde{\text{T}}$ eigenbasis~\eqref{eq:numberCPT}, the transport equations~\eqref{eq:dotnN}--\eqref{eq:dotnbarE} for the $N$-$L$-$E$ thermal system read
\begin{eqnarray}
	\label{eq:dotNeven}
	\dot{\underline{n}}^N &=& -\, \frac{i}{2}\,\Big[\,{\cal E}_N\,, \Delta n^N\Big]\: -\: 
	\frac{1}{2n^N_{\rm eq}}\,
	\Big\{ \underline{n}^N\,, \text{Tr}_L\big(\underline{\gamma}^N_{L\Phi}\big)\Big\}\:
	-\: 
	\frac{1}{2n^N_{\rm eq}}\,
	\Big\{ \Delta n^N\,, \text{Tr}_L\big(\delta\gamma^N_{L\Phi}\big)\Big\}\:
	+\: 
	\frac{1}{n^L_{\rm eq}}\, \text{Tr}_{L}\big(\underline{n}^L\, \underline{\gamma}^{L\Phi}_N\big)\nonumber\\
	&&+\: \frac{1}{n^L_{\rm eq}}\, \text{Tr}_{L}\big(\Delta n^L\, \delta\gamma^{L\Phi}_N\big)\: -\: \frac{1}{2n^N_{\rm eq}n^L_{\rm eq}}\, \Big\{ \underline{n}^N\,, \text{Tr}_L\big(
	\underline{\gamma}^{\bar{N}L}_{\bar\Phi}, \underline{n}^L\big)\Big\}\:
	+\: \frac{1}{2n^N_{\rm eq}n^L_{\rm eq}}\, \Big\{ \Delta n^N\,, \text{Tr}_L\big(
	\underline{\gamma}^{\bar{N}L}_{\bar\Phi}\, \Delta n^L\big)\Big\}\nonumber\\
	&&+\: \frac{1}{2n^N_{\rm eq}n^L_{\rm eq}}\, \Big\{ \Delta n^N\,, \text{Tr}_L\big(\delta\gamma^{\bar{N}L}_{\bar\Phi}
	\underline{n}^L
	\big)\Big\}\: -\: \frac{1}{2n^N_{\rm eq}n^L_{\rm eq}}\, \Big\{ \underline{n}^N\,, \text{Tr}_L\big(\delta\gamma^{\bar{N}L}_{\bar\Phi}\Delta n^L\big)\Big\}\: +\:  \text{Tr}_L\big(
	\underline{\gamma}_{\bar{N}L}^{\bar\Phi}\big)\,,\qquad
	\\[20mm]
	\label{eq:dotNodd}
	\Delta \dot{n}^N &=& -\, 2i\,\Big[\,{\cal E}_N\,, \underline{n}^N\Big]\: 
	-\: 
	\frac{1}{2n^N_{\rm eq}}\,
	\Big\{ \underline{n}^N\,, \text{Tr}_L\big(\delta\gamma^N_{L\Phi}\big)\Big\}\:
	-\: 
	\frac{1}{2n^N_{\rm eq}}\,
	\Big\{ \Delta n^N\,, \text{Tr}_L\big(\underline{\gamma}^N_{L\Phi}\big)\Big\}\:
	\boldsymbol{\color{blue}{-}}\: 
	\frac{1}{n^L_{\rm eq}}\, \text{Tr}_{L}\big(\underline{n}^L\, \delta\gamma^{L\Phi}_N\big)\nonumber\\
	&&+\: \frac{1}{n^L_{\rm eq}}\, \text{Tr}_{L}\big(\Delta n^L\, \underline{\gamma}^{L\Phi}_N\big)\: +\: \frac{1}{2n^N_{\rm eq}n^L_{\rm eq}}\, \Big\{ \underline{n}^N\,, \text{Tr}_L\big(\delta\gamma^{\bar{N}L}_{\bar\Phi}\, \underline{n}^L\big)\Big\}\:
	-\: \frac{1}{2n^N_{\rm eq}n^L_{\rm eq}}\, \Big\{ \Delta n^N\,, \text{Tr}_L\big(\delta\gamma^{\bar{N}L}_{\bar\Phi} \Delta n^L\big)\Big\}\nonumber\\
	&&-\: \frac{1}{2n^N_{\rm eq}n^L_{\rm eq}}\, \Big\{ \Delta n^N\,, \text{Tr}_L\big(
	\underline{\gamma}^{\bar{N}L}_{\bar\Phi}\, \underline{n}^L\big)\Big\}\: +\: \frac{1}{2n^N_{\rm eq}n^L_{\rm eq}}\, \Big\{ \underline{n}^N\,, \text{Tr}_L\big(
	\underline{\gamma}^{\bar{N}L}_{\bar\Phi} \Delta n^L\big)\Big\}\: \boldsymbol{\color{blue}{+}}\:  \text{Tr}_L\big(\delta
	\underline{\gamma}^{\bar{N}L}_{\bar\Phi} \big)\,,
	\\[5mm]
	\label{eq:dotLeven}
	\dot{\underline{n}}^L &=& -\, \frac{i}{2}\,\Big[\,{\cal E}_L\,, \Delta n^L\Big]\: -\: 
	\frac{1}{2n^L_{\rm eq}}\,
	\Big\{ \underline{n}^L\,, \text{Tr}_N\big(\underline{\gamma}_N^{L\Phi}\big)\Big\}
	-\: 
	\frac{1}{2n^L_{\rm eq}}\,
	\Big\{ \Delta n^L\,, \text{Tr}_N\big(\delta\gamma_N^{L\Phi}\big)\Big\}
	\: +\: \frac{1}{n^N_{\rm eq}}\, \text{Tr}_{N}\big(\underline{n}^N \underline{\gamma}_{L\Phi}^N\big)\nonumber\\
	&&+\: 
	\frac{1}{n^N_{\rm eq}}\, \text{Tr}_{N}\big(\Delta n^N \delta\gamma_{L\Phi}^N\big)\: 
	-\: \frac{1}{2n^N_{\rm eq}n^L_{\rm eq}}\,
	\Big\{ \underline{n}^L\,, \text{Tr}_{\bar{N}}\big(\underline{n}^N \underline{\gamma}^{\bar{N}L}_{\bar{\Phi}} \big)\Big\}\: 
	+\: \frac{1}{2n^N_{\rm eq}n^L_{\rm eq}}\,
	\Big\{ \Delta n^L\,, \text{Tr}_{\bar{N}}\big(\Delta n^N \underline{\gamma}^{\bar{N}L}_{\bar{\Phi}} \big)\Big\}\nonumber\\
	&&-\: \frac{1}{2n^N_{\rm eq}n^L_{\rm eq}}\,
	\Big\{ \Delta n^L\,, \text{Tr}_{\bar{N}}\big(\underline{n}^N \delta\gamma^{\bar{N}L}_{\bar{\Phi}} \big)\Big\}\: 
	+\: \frac{1}{2n^N_{\rm eq}n^L_{\rm eq}}\,
	\Big\{ \underline{n}^L\,, \text{Tr}_{\bar{N}}\big(\Delta n^N \delta\gamma^{\bar{N}L}_{\bar{\Phi}} \big)\Big\}\:
	+\:  \text{Tr}_{\bar{N}}\big(\underline{\gamma}^{\bar{\Phi}}_{\bar{N}L}\big)\nonumber\\
	&&-\: \frac{1}{2n^L_{\rm eq}n^E_{\rm eq}}\,
	\Big\{ \underline{n}^L\,, \text{Tr}_{\bar{E}}\big(\underline{n}^E \underline{\gamma}^{\bar{E}L}_{\bar{\Phi}} \big)\Big\}\: +\:  
	\frac{1}{2n^L_{\rm eq}n^E_{\rm eq}}\,
	\Big\{ \Delta n^L\,, \text{Tr}_{\bar{E}}\big(\Delta n^E \underline{\gamma}^{\bar{E}L}_{\bar{\Phi}} \big)\Big\}\nonumber\\
	&& -\:
	\frac{1}{2n^L_{\rm eq}n^E_{\rm eq}}\,
	\Big\{ \Delta n^L\,, \text{Tr}_{\bar{E}}\big(\bar{n}^E \delta\gamma^{\bar{E}L}_{\bar{\Phi}} \big)\Big\}\: +\:
	\frac{1}{2n^L_{\rm eq}n^E_{\rm eq}}\,
	\Big\{ \underline{n}^L\,, \text{Tr}_{\bar{E}}\big(\Delta n^E \delta\gamma^{\bar{E}L}_{\bar{\Phi}} \big)\Big\}\: +\:
	\text{Tr}_{\bar{E}}\big(\underline{\gamma}^{\bar{\Phi}}_{\bar{E}L}\big)\nonumber\\
	&&-\: \frac{1}{2n^L_{\rm eq}}\,
	\Big\{ \underline{n}^L\,, \text{Tr}_{\bar{L}}\big(\underline{\gamma}^{L\bar{L}}_{W_\mu} \underline{n}^L \big)\Big\}\: +\:  
	\frac{1}{2n^L_{\rm eq}}\,
	\Big\{ \Delta n^L\,, \text{Tr}_{\bar{L}}\big(\underline{\gamma}^{L\bar{L}}_{W_\mu} \Delta n^L \big)\Big\}\: +\:
	\text{Tr}_{\bar{L}}\big(\underline{\gamma}^{W_\mu}_{L\bar{L}}\big)\nonumber\\
	&&-\: \frac{1}{2n^L_{\rm eq}}\,
	\Big\{ \Delta n^L\,, \text{Tr}_{\bar{L}}\big(\delta\gamma^{L\bar{L}}_{W_\mu} \underline{n}^L \big)\Big\}\: +\:  
	\frac{1}{2n^L_{\rm eq}}\,
	\Big\{ \underline{n}^L\,, \text{Tr}_{\bar{L}}\big(\delta\gamma^{L\bar{L}}_{W_\mu} \Delta n^L \big)\Big\}\,,
	\\[5mm]
	\label{eq:dotLodd}
	\Delta \dot{n}^L &=& -\, 2i\,\Big[\,{\cal E}_L\,, \underline{n}^L\Big]\: 
	\boldsymbol{\color{blue}{+}}\: 
	\frac{1}{2n^L_{\rm eq}}\,
	\Big\{ \underline{n}^L\,, \text{Tr}_N\big(\delta\gamma_N^{L\Phi}\big)\Big\}
	-\: 
	\frac{1}{2n^L_{\rm eq}}\,
	\Big\{ \Delta n^L\,, \text{Tr}_N\big(\underline{\gamma}_N^{L\Phi}\big)\Big\}
	\: +\: \frac{1}{n^N_{\rm eq}}\, \text{Tr}_{N}\big(\Delta n^N \underline{\gamma}_{L\Phi}^N\big)\nonumber\\
	&&+\: 
	\frac{1}{n^N_{\rm eq}}\, \text{Tr}_{N}\big(\underline{n}^N \delta\gamma_{L\Phi}^N\big)\: 
	-\: \frac{1}{2n^N_{\rm eq}n^L_{\rm eq}}\,
	\Big\{ \underline{n}^L\,, \text{Tr}_{\bar{N}}\big(\underline{n}^N \delta\gamma^{\bar{N}L}_{\bar{\Phi}} \big)\Big\}\: 
	+\: \frac{1}{2n^N_{\rm eq}n^L_{\rm eq}}\,
	\Big\{ \Delta n^L\,, \text{Tr}_{\bar{N}}\big(\Delta n^N \delta\gamma^{\bar{N}L}_{\bar{\Phi}} \big)\Big\}\nonumber\\
	&&-\: \frac{1}{2n^N_{\rm eq}n^L_{\rm eq}}\,
	\Big\{ \Delta n^L\,, \text{Tr}_{\bar{N}}\big(\underline{n}^N \underline{\gamma}^{\bar{N}L}_{\bar{\Phi}} \big)\Big\}\: 
	+\: \frac{1}{2n^N_{\rm eq}n^L_{\rm eq}}\,
	\Big\{ \underline{n}^L\,, \text{Tr}_{\bar{N}}\big(\Delta n^N \underline{\gamma}^{\bar{N}L}_{\bar{\Phi}} \big)\Big\}\:
	\boldsymbol{\color{blue}{-}}\: \text{Tr}_{\bar{N}}\big(\delta\gamma^{\bar{\Phi}}_{\bar{N}L}\big)\nonumber\\
	&&-\: \frac{1}{2n^L_{\rm eq}n^E_{\rm eq}}\,
	\Big\{ \underline{n}^L\,, \text{Tr}_{\bar{E}}\big(\underline{n}^E \delta\gamma^{\bar{E}L}_{\bar{\Phi}} \big)\Big\}\: +\:  
	\frac{1}{2n^L_{\rm eq}n^E_{\rm eq}}\,
	\Big\{ \Delta n^L\,, \text{Tr}_{\bar{E}}\big(\Delta n^E \delta\gamma^{\bar{E}L}_{\bar{\Phi}} \big)\Big\}\nonumber\\
	&& -\:
	\frac{1}{2n^L_{\rm eq}n^E_{\rm eq}}\,
	\Big\{ \Delta n^L\,, \text{Tr}_{\bar{E}}\big(\bar{n}^E \underline{\gamma}^{\bar{E}L}_{\bar{\Phi}} \big)\Big\}\: +\:
	\frac{1}{2n^L_{\rm eq}n^E_{\rm eq}}\,
	\Big\{ \underline{n}^L\,, \text{Tr}_{\bar{E}}\big(\Delta n^E \underline{\gamma}^{\bar{E}L}_{\bar{\Phi}} \big)\Big\}\: 
	\boldsymbol{\color{blue}{-}}\:
	\text{Tr}_{\bar{E}}\big(\delta\gamma^{\bar{\Phi}}_{\bar{E}L}\big)\nonumber\\
	&&-\: \frac{1}{2n^L_{\rm eq}}\,
	\Big\{ \Delta n^L\,, \text{Tr}_{\bar{L}}\big(\underline{\gamma}^{L\bar{L}}_{W_\mu} \underline{n}^L \big)\Big\}\: +\:  
	\frac{1}{2n^L_{\rm eq}}\,
	\Big\{ \underline{n}^L\,, \text{Tr}_{\bar{L}}\big(\underline{\gamma}^{L\bar{L}}_{W_\mu} \Delta n^L \big)\Big\}\nonumber\\
	&&-\: \frac{1}{2n^L_{\rm eq}}\,
	\Big\{ \underline{n}^L\,, \text{Tr}_{\bar{L}}\big(\delta\gamma^{L\bar{L}}_{W_\mu} \underline{n}^L \big)\Big\}\: +\:  
	\frac{1}{2n^L_{\rm eq}}\,
	\Big\{ \Delta n^L\,, \text{Tr}_{\bar{L}}\big(\delta\gamma^{L\bar{L}}_{W_\mu} \Delta n^L \big)\Big\}\: \boldsymbol{\color{blue}{-}}\:
	\text{Tr}_{\bar{L}}\big(\delta\gamma^{W_\mu}_{L\bar{L}}\big)\,,\\[5mm]
	\label{eq:dotEeven}
	\dot{\underline{n}}^E &=& -\, \frac{i}{2}\,\Big[\,{\cal E}_E\,, \Delta n^E \Big]\:
	-\: \frac{1}{2n^E_{\rm eq}n^L_{\rm eq}}\,
	\Big\{ \underline{n}^E\,, \text{Tr}_L\big(
	\underline{\gamma}^{\bar{E}L}_{\Phi} \underline{n}^L\big)\Big\}
	\: +\: 
	\frac{1}{2n^E_{\rm eq}n^L_{\rm eq}}\,
	\Big\{ \Delta n^E\,, \text{Tr}_L\big(
	\underline{\gamma}^{\bar{E}L}_{\Phi} \Delta n^L\big)\Big\}\nonumber\\
	&&+\:
	\frac{1}{2n^E_{\rm eq}n^L_{\rm eq}}\,
	\Big\{ \Delta n^E\,, \text{Tr}_L\big(\delta\underline{\gamma}^{\bar{E}L}_{\Phi} \underline{n}^L\big)\Big\}\: -\:
	\frac{1}{2n^E_{\rm eq}n^L_{\rm eq}}\,
	\Big\{ \underline{n}^E\,, \text{Tr}_L\big(
	\delta\gamma^{\bar{E}L}_{\Phi} \Delta n^L\big)\Big\}\: +\:
	\text{Tr}_L\big(
	\underline{\gamma}_{\bar{E}L}^{\Phi}\big)\nonumber\\
	&&-\: \frac{1}{2n^E_{\rm eq}}\,
	\Big\{ \underline{n}^E\,, \text{Tr}_{\bar{E}}\big( \underline{\gamma}^{E\bar{E}}_{B_\mu}\, \underline{n}^E\big)\Big\}\: 
	+\:  \frac{1}{2n^E_{\rm eq}}\,
	\Big\{ \Delta n^E\,, \text{Tr}_{\bar{E}}\big( \underline{\gamma}^{E\bar{E}}_{B_\mu}\, \Delta n^E\big)\Big\}\: +\: 
	\text{Tr}_{\bar{E}}\big(\underline{\gamma}^{B_\mu}_{E\bar{E}}\big)
	\nonumber\\
	&&-\: \frac{1}{2n^L_{\rm eq}}\,
	\Big\{ \Delta n^E\,, \text{Tr}_{\bar{E}}\big(\delta\gamma^{E\bar{E}}_{B_\mu} \underline{n}^E \big)\Big\}\: +\:  
	\frac{1}{2n^L_{\rm eq}}\,
	\Big\{ \underline{n}^E\,, \text{Tr}_{\bar{E}}\big(\delta\gamma^{E\bar{E}}_{B_\mu} \Delta n^E \big)\Big\}\,,\\[20mm]
	\label{eq:dotEodd}
	\Delta \dot{n}^E &=& -\, 2i\,\Big[\,{\cal E}_E\,, \underline{n}^E\Big]\: 
	+\: \frac{1}{2n^E_{\rm eq}n^L_{\rm eq}}\,
	\Big\{ \underline{n}^E\,, \text{Tr}_L\big(\delta\gamma^{\bar{E}L}_{\Phi} \underline{n}^L\big)\Big\}
	\: -\: 
	\frac{1}{2n^E_{\rm eq}n^L_{\rm eq}}\,
	\Big\{ \Delta n^E\,, \text{Tr}_L\big(\delta\gamma^{\bar{E}L}_{\Phi} \Delta n^L\big)\Big\}\qquad\qquad\nonumber\\
	&&-\:
	\frac{1}{2n^E_{\rm eq}n^L_{\rm eq}}\,
	\Big\{ \Delta n^E\,, \text{Tr}_L\big(
	\underline{\gamma}^{\bar{E}L}_{\Phi} \underline{n}^L\big)\Big\}\: +\:
	\frac{1}{2n^E_{\rm eq}n^L_{\rm eq}}\,
	\Big\{ \underline{n}^E\,, \text{Tr}_L\big(
	\underline{\gamma}^{\bar{E}L}_{\Phi} \Delta n^L\big)\Big\}\: \boldsymbol{\color{blue}{+}}\:
	\text{Tr}_L\big(\delta\gamma_{\bar{E}L}^{\Phi}\big)\nonumber\\
	&&-\: \frac{1}{2n^L_{\rm eq}}\,
	\Big\{ \Delta n^E\,, \text{Tr}_{\bar{E}}\big(\underline{\gamma}^{E\bar{E}}_{B_\mu} \underline{n}^E \big)\Big\}\: +\:  
	\frac{1}{2n^L_{\rm eq}}\,
	\Big\{ \underline{n}^E\,, \text{Tr}_{\bar{E}}\big(\underline{\gamma}^{E\bar{E}}_{B_\mu} \Delta n^E \big)\Big\}\nonumber\\
	&&-\: \frac{1}{2n^E_{\rm eq}}\,
	\Big\{ \underline{n}^E\,, \text{Tr}_{\bar{E}}\big( \delta\gamma^{E\bar{E}}_{B_\mu}\, \underline{n}^E\big)\Big\}\: 
	+\:  \frac{1}{2n^E_{\rm eq}}\,
	\Big\{ \Delta n^E\,, \text{Tr}_{\bar{E}}\big( \delta\gamma^{E\bar{E}}_{B_\mu}\, \Delta n^E\big)\Big\}\: 
	\boldsymbol{\color{blue}{-}}\:  
	\text{Tr}_{\bar{E}}\big(\delta\gamma^{B_\mu}_{E\bar{E}}\big)
	\, .
\end{eqnarray}
In \eqref{eq:dotNodd}, \eqref{eq:dotLodd} and \eqref{eq:dotEodd}, we included contributions due to the so-called real intermediate states (RIS) from CP-violating $2 \leftrightarrow 2$ scatterings that changed the sign of some collision terms, so as to ensure that no net asymmetries $\Delta n^{N,L,E}$ are generated in thermal equilibrium, e.g.~see~\cite{BhupalDev:2014pfm}. These sign flips due to CP-odd RIS are indicated in boldface and in blue. It is important to stress again that  \eqref{eq:dotNeven}--\eqref{eq:dotEodd} provide an effective yet-thermally complete description of the flavour-covariant transport equations for all relevant $1\leftrightarrow 2$ rates that can take place in the entire $N$-$L$-$E$ system under the product flavour group~$G_{\rm F}$ given in~\eqref{eq:Gflavour} to all orders.

It is instructive to verify that, in the absence of the $B+L$-violating sphaleron interactions,  the total lepton number ${L_{\rm tot} = L_{\rm SM} + L_N}$ (with $L_{\rm SM}\equiv L_L + L_E$) is preserved by the transport equations stated in~\eqref{eq:dotNeven}--\eqref{eq:dotEodd}, when the heavy neutrinos $N_i$ are Dirac particles. To this end, we must prove that the equality
\begin{equation}
	\label{eq:Ltot} 
	\frac{\rm d}{{\rm d}t}\, \Big(\text{Tr}_N\big(\Delta n^N\big)\: +\:  \text{Tr}_L\big(\Delta n^L\big)\: +\:  \text{Tr}_E\big(\Delta n^E\big)\Big)\ =\ 0\,,  
\end{equation}
holds for all cosmic times $t$. Although proving~\eqref{eq:Ltot} may seem tedious, there are a few properties among the various collision terms that facilitate this task. First, observe that all commutators $\big[{\cal E}_X\,, \underline{n}^X\big]$ (with $X=N,L,E$) vanish when taking the trace in the flavour space. Second, one can find numerous cancellations between terms by employing identities, such as 
\begin{eqnarray}
	\label{eq:TrXid}
	\text{Tr}_{X}\Big\{ \underline{n}^X\,, \text{Tr}_Y\big(\delta\gamma^X_{Y\Phi}\big)\Big\}\!&=&\! 
	2\,\text{Tr}_{X,Y}\Big(\underline{n}^X\,\delta\gamma^X_{Y\Phi}\Big)\,,\\
	\label{eq:TrXYid}
	\text{Tr}_{X}\Big\{ \underline{n}^X\,, \text{Tr}_{\bar{Y}}\big(\delta\gamma^{X\bar{Y}}_{\Phi}\underline{n}^Y\big)\Big\}\!&=&\! 
	\text{Tr}_{X,\bar{Y}}\Big(\delta\gamma^{X\bar{Y}}_{\Phi} \big\{ \underline{n}^X\,, \underline{n}^Y\big\}\Big)\: =\: \text{Tr}_{\bar{X},Y}\Big(\delta\gamma_{\bar{X}Y}^{\bar{\Phi}} \big\{ \underline{n}^X\,, \underline{n}^Y\big\}\Big)\,,
\end{eqnarray}
with $X,Y\in \{N,L,E\}$. Finally, it is not difficult to convince ourselves that the gauge interactions do not affect the lepton flavour coherences $\Delta n^L$ and $\Delta n^E$ given in~\eqref{eq:dotLodd} and~\eqref{eq:dotEodd}, respectively.
It can be explicitly checked that the collision terms associated with the $W^{0,1,2,3}_\mu$ gauge bosons cancel identically in both $\text{Tr}_L\big(\Delta n^L\big)$ and $\text{Tr}_E\big(\Delta n^E\big)$, by virtue of 
\begin{equation}
	\text{Tr}_X\Big\{ \Delta n^X\,,\text{Tr}_{\bar{X}}\big(\underline{\gamma}^{X\bar{X}}_{W_\mu} \underline{n}^X \big)\Big\}\: =\:  
	\text{Tr}_X\Big\{ \underline{n}^X\,, \text{Tr}_{\bar{X}}\big(\underline{\gamma}^{X\bar{X}}_{W_\mu}\,\Delta n^X \big)\Big\}\: =\:  \text{Tr}_{X,\bar{X}}\Big(\underline{\gamma}^{X\bar{X}}_{W_\mu} \big\{ \underline{n}^X\,,\Delta n^X \big\}\Big)\, .  
\end{equation}
Moreover, thanks to the self-adjoint structure of $\gamma_{X\bar{X}}^{W_\mu}$ in \eqref{eq:gamWmu}, 
it is easy to show that the CP-violating gauge-mediated collision terms, such as $\delta\gamma_{X\bar{X}}^{W_\mu}$, do not contribute to the transport equations after taking the trace over both the $X$ and $\bar{X}$ flavour spaces, i.e.~$\text{Tr}_{X,\bar{X}}\big(\delta\gamma^{W_\mu}_{X\bar{X}}\big) = 0$ and $\text{Tr}_{X,\bar{X}}\big(\underline{n}^X\delta\gamma^{W_\mu}_{X\bar{X}}\,\underline{n}^X\big) = \text{Tr}_{X,\bar{X}}\big(\Delta n^X\delta\gamma^{W_\mu}_{X\bar{X}}\,\Delta n^X\big) = 0$. 
These findings are consistent with those derived using the Kadanoff-Baym formalism in~\cite{Li:2026qym}, where flavour-universal gauge interactions that are invariant under the action of the flavour group $G_{\rm F}$ will not destroy the net effect of quantum coherences in $\Delta n^L$ and $\Delta n^E$.

\vfill\eject

\section{Numerical analysis with Markov Chain Monte Carlo and Deep Neural Network}

\subsection{Setup for parameter scan}

Throughout the numerical analysis based on  biased and unbiased Markov Chain Monte Carlo (MCMC) scan, as well as the further Deep Neural Network (DNN) training of the MCMC data, we adopt the Casas-Ibarra (CI) parametrization~\cite{Casas:2001sr} to perform the numerical evaluation with the full three singlet  neutrino generations,
\begin{align}
	h_{\alpha i}=\frac{i\sqrt{2}}{v}\sum_{\beta}U_{\nu,\alpha \beta}\sqrt{m_\beta}\:\mathbb{R}_{\beta i}\sqrt{M_i}\,,
\end{align}
where  $v\approx 246$~GeV is the electroweak vacuum expectation value, $U_{\nu,\alpha\beta}$ with $\alpha,\beta=e,\mu,\tau$ denotes the $3\times 3$ Pontecorvo-Maki-Nakagawa-Sakata (PMNS) matrix in the standard parametrization of~\cite{ParticleDataGroup:2024cfk}, and $i$ denotes the singlet neutrino flavour. The full $\mathbb R$ matrix entering the CI formula is the ordered product
\begin{equation}
	\mathbb{R}=
	R_{23}(\omega_{23})\, R_{13}(\omega_{13})\, R_{12}(\omega_{12})\,,
\end{equation}
which is itself complex orthogonal, $\mathbb R^T \mathbb R = \mathbb{I}$. The complex rotation in the $(i,j)$ plane reads
\begin{equation}
	R_{12}(\omega) =
	\begin{pmatrix}
		\cos\omega_{12} & \sin\omega_{12} & 0 \\
		-\sin\omega_{12} & \cos\omega_{12} & 0 \\
		0 & 0 & 1
	\end{pmatrix}\!,\quad
	R_{13}(\omega) =
	\begin{pmatrix}
		\cos\omega_{13} & 0 & \sin\omega_{13} \\
		0 & 1 & 0 \\
		-\sin\omega_{13} & 0 & \cos\omega_{13}
	\end{pmatrix}\!,\quad
	R_{23}(\omega) =
	\begin{pmatrix}
		1 & 0 & 0 \\
		0 & \cos\omega_{23} & \sin\omega_{23} \\
		0 & -\sin\omega_{23} & \cos\omega_{23}
	\end{pmatrix}\,,
\end{equation}
with  the complex rotation angle $\omega_{ij}$ parametrized by
\begin{align}
	\omega_{12} = \mathrm{Re}(\omega_{12}) + i\,\mathrm{Im}(\omega_{12})\,,\qquad
	\omega_{13} = \mathrm{Re}(\omega_{13}) + i\,\mathrm{Im}(\omega_{13})\,,\qquad
	\omega_{23} = \mathrm{Re}(\omega_{23}) + i\,\mathrm{Im}(\omega_{23})\,.
\end{align}

\begin{table}[h]
	\centering
	\begin{tabular}{ccc}
		\toprule
		\textbf{Parameter} & \textbf{Symbol} & \textbf{Range / value} \\
		\midrule
		Dirac CP phase        & $\delta_\text{CP}$       & $[0,\, 2\pi]$ \\
		Majorana phase 1      & $\alpha_1$               & $[0,\, 2\pi]$ \\
		Majorana phase 2      & $\alpha_2$               & $[0,\, 2\pi]$ \\
		Real $R_{12}$ & $\mathrm{Re}(\omega_{12})$ & $[0,\, 2\pi]$ \\
		Real $R_{13}$ & $\mathrm{Re}(\omega_{13})$ & $[0,\, 2\pi]$ \\
		Real $R_{23}$ & $\mathrm{Re}(\omega_{23})$ & $[0,\, 2\pi]$ \\
		Imaginary $R_{12}$ & $\mathrm{Im}(\omega_{12})$ & $[-10,\, 10]$ \\
		Imaginary $R_{13}$ & $\mathrm{Im}(\omega_{13})$ & $[-10,\, 10]$ \\
		Imaginary $R_{23}$ & $\mathrm{Im}(\omega_{23})$ & $[-10,\, 10]$ \\
		Heavy neutrino mass 1 & $M_1$                    & $[1,\, 50]$~GeV \\
		Heavy neutrino mass 2 & $M_2$                    & $[1,\, 50]$~GeV \\
		Lightest neutrino mass & $m_1(\text{NO}),m_3(\text{IO})$                   & $0.01$~eV \text{or} $0$~eV \\
		\bottomrule
		\toprule
	\end{tabular}
	\caption{Scan ranges for PMNS phases, the CI parameters, singlet neutrino masses, as well as the lightest active neutrino mass in the NO and IO case. No ordering between $M_1$ and $M_2$ is imposed.}
	\label{tab:CI_params}
\end{table}

In evaluating the contribution to the BAU, we consider two GeV-scale sterile neutrinos $N_1, N_2$, and decouple the third one $N_3$. Effectively, there are two ways to decouple $N_3$.   In the first case, $N_3$ is lighter than the other two. Since $h_{\alpha i}\propto \sqrt{M_i}$ under the CI parametrization, the associated Yukawa couplings $h_{\alpha 3}$ are much smaller than $h_{\alpha 2}, h_{\alpha 1}$: $h_{\alpha 3}\ll h_{\alpha 2}, h_{\alpha 1}$, physically indicating that only two neutrino singlets contribute significantly to the BAU. In the second case, $N_3$ is much heavier, and the associated lepton asymmetry is generated much earlier in the Universe by $N_3$ decay. However, before the Universe cools down to the freeze-in temperatures of the $N_1,N_2$ production, the lepton asymmetry generated from the $N_3$ decay already underwent  washout in lepton-number breaking processes. Additionally, at later times, Higgs decay to $N_3$ is kinematically forbidden and hence does not contribute to the BAU. Here, we adopt the first decoupling case by considering a much lighter $N_3$, $M_3\ll M_1, M_2$. 

The ranges of scan for the PMNS Dirac CP phase, Majorana phases, the CI parameters, as well as the sterile neutrino masses $M_1,M_2$ are shown in Tab.~\ref{tab:CI_params}. For normal ordering (NO) or inverted ordering (IO)  of the active neutrino mass spectrum,  we consider two benchmark scenarios for the lightest active neutrino mass, i.e., a non-zero mass: $m_1=0.01~(m_3=0.01)$~eV for NO (IO), and a vanishing mass: $m_1=0~(m_3=0)$~eV for NO (IO).

\subsection{Scan routine: unbiased MCMC$\to$ biased MCMC$\to$DNN training.}
We first perform an unbiased MCMC scan with $10^6$ sample points and select those that yield the correct order of the BAU at $Y_B=[10^{-11},10^{-10}]$, where the observed value $Y_B\approx 8.75\times 10^{-11}$ resides. The unbiased scan does not impose any preferred cut on  the free CI parameters or on  the Yukawa matrix elements $h_{\alpha i}$. In this unbiased scan, the lightest SM neutrino mass is not fixed but varies in the range of $[10^{-5},0.03]$~eV. 

For the NO case, of the $10^6$ scanned points, only $6454$ land in the
correct-order window, corresponding to a hit-rate of $0.65\%$. It suggests that reproducing the observed baryon asymmetry is, if possible, numerically challenging without any preferred cut on scan over the totally 12 free parameters. We document the distribution of the Yukawa coupling ratio $|h_e/h_\mu|=\sqrt{\sum_i |h_{ei}|^2}/\sqrt{\sum_j |h_{\mu j}|^2}$ in Tab.~\ref{tab:unbiased_ratio_NO}
among those $6454$ points. Since the Yukawa coupling ratio $|h_e|/|h_\mu|$ characterizes the relative washout effect between the electron and muon flavour asymmetries, a large ratio implies that a hierarchical washout pattern is favoured in TIRL to generate the required BAU.  In the NO case, Tab.~\ref{tab:unbiased_ratio_NO} shows that stronger washout in the muon flavour asymmetry is preferred. 

\begin{table}[h!]
	\centering
	\begin{tabular}{cccc}
		\toprule
		$|h_e|/|h_\mu|$ bin & No. of points  & \% of all scanned & \% of correct $Y_B$ \\
		\midrule
		$< 0.01$            & 0    & 0.0000\%  & 0.0000\%  \\
		$0.01$--$0.1$       & 6268 & 0.6268\%  & \textbf{97.1181\%} \\
		$0.1$--$1$          & 163  & 0.0163\%  & 2.5256\%  \\
		$1$--$10$           & 23   & 0.0023\%  & 0.3564\%  \\
		$10$--$100$         & 0    & 0.0000\%  & 0.0000\%  \\
		$>100$              & 0    & 0.0000\%  & 0.0000\%  \\
		\midrule
		Total               & 6454 & 0.6454\%  & 100.0000\% \\
		\bottomrule
		\toprule
	\end{tabular}
	\caption{Statistic distribution of $|h_e|/|h_\mu|=\sqrt{\sum_i |h_{ei}|^2}/\sqrt{\sum_i|h_{\mu i}|^2}$ among the $6454$ (of $10^6$)
		unbiased MCMC scan points hitting the correct-order $Y_B$ for the NO case. The rightmost  column is the  counts renormalized only
		over the correct-$Y_B$ subset, i.e., the conditional distribution of the ratio given correct $Y_B=[10^{-11},10^{-10}]$.}
	\label{tab:unbiased_ratio_NO}
\end{table}

For the IO case, 
of the $10^6$ scanned points, $17913$  land in the correct-order
window of $Y_B$, corresponding to a hit-rate of $1.8\%$. Table~\ref{tab:unbiased_ratio_IO} gives the full breakdown of
$|h_e|/|h_\mu|$ among  those $17913$ points, where a larger washout effect on the electron flavour asymmetry is instead favoured, comparing to the NO case. 

\begin{table}[h!]
	\centering
	\begin{tabular}{lrrr}
		\toprule
		$|h_e|/|h_\mu|$ bin & No. of points & \% of all scanned & \% of correct-$Y_B$ \\
		\midrule
		$< 0.01$            & 5     & 0.0005\%  & 0.0279\%  \\
		$0.01$--$0.1$       & 1282  & 0.1282\%  & 7.1568\%  \\
		$0.1$--$1$          & 165   & 0.0165\%  & 0.9211\%  \\
		$1$--$10$           & 167   & 0.0167\%  & 0.9323\%  \\
		$10$--$100$         & 16294 & 1.6294\%  & \textbf{90.9619\%} \\
		$>100$              & 0     & 0.0000\%  & 0.0000\%  \\
		\midrule
		Total               & 17913 & 1.7913\%  & 100.0000\% \\
		\bottomrule
		\toprule
	\end{tabular}
	\caption{Statistic distribution of $|h_e|/|h_\mu|$ among the $17913$ (of $10^6$)
		unbiased MCMC scan points hitting the correct-order $Y_B$ for the IO case.}
	\label{tab:unbiased_ratio_IO}
\end{table}

The distribution of $|h_e|/|h_\mu|$ for both the NO and IO cases provide a useful guide to find out more available parameter space in a more efficient way under the multi-dimensional parameter scan. Following the different hierarchical washout patterns in NO and IO, we next perform a biased MCMC scan by fixing two benchmark values for the lightest neutrino mass, $0.01$~eV and 0~eV. We impose the cut conditions (hierarchy prior),
\begin{align}
	\text{NO prior (muon dominance)}:~\frac{|h_\mu|}{|h_e|}>10\,,\quad   \text{IO prior (electron dominance)}:~\frac{|h_e|}{|h_\mu|}>10\,.
\end{align}
Of the $10^6$ samples in the NO prior, we indeed find that the successful points increase to a hit-rate of about $40\%$ for $m_1=0.01$~eV and to a hit-rate of about $17\%$ for $m_1=0$~eV, comparing to the unbiased MCMC  hit-rate $0.65\%$. For the IO prior under the $10^6$ sample points, on the other hand, we find that both the hit-rates for $m_3=0.01$~eV and $m_3=0$~eV increase to about $40\%$ from the unbiased rate $1.8\%$.

The parameter space of the CI parametrization is 12-dimensional
(11 sampled parameters plus the fixed lightest neutrino mass $m_1$ or $m_3$). Direct grid scanning is computationally intractable,
while standard MCMC methods can suffer from slow mixing and poor coverage of disconnected high-probability regions. A normalizing flow overcomes these limitations by learning an invertible, differentiable map between a simple base distribution (multivariate Gaussian) and the complex target distribution
defined by the MCMC posterior. To show the general parameter space that predicts $Y_B=[10^{-11},10^{-10}]$, we hence conduct a final stage by training the biased MCMC data under the approach of DNN normalizing flow based on the real-valued non-volume preserving transformations~\cite{Dinh:2016q}. Once trained, the flow generates $10^6$ new candidate parameter vectors by sampling the base Gaussian and pushing the samples through the learned invertible map, allowing the DNN stage to fill in and smooth the MCMC posterior far more densely than continuing the MCMC chain itself could achieve. We should emphasize that  every generated sample must be revalidated by solving the kinetic equations of lepton asymmetry, and then filtered again to the correct order of the $Y_B$ window for plotting.

\subsection{More explanations of the parameter space}

After the final stage of the DDN training, we plot the main figures in terms of the following variable set
\begin{align}\label{eq:density_plot}
	\left(M_1, \theta_\alpha^2, \frac{M_2-M_1}{M_1}\equiv \frac{\Delta M}{M_1}\right)\,,
\end{align}
where $\theta_\alpha$ denotes the mixing angle between the SM neutrinos and the heavy neutrino singlets, with
\begin{align}
	\theta_\alpha^2\equiv \sum_i\theta_{\alpha i}^2\equiv\sum_{i} \frac{v^2|h_{\alpha i}|^2}{M^2_i}\,.
\end{align}
The mixing angle $\theta_\alpha$ characterizes the interaction between the GeV-scale singlet neutrinos and SM particles, which is an observable parameter in experimental searches. In a more general consideration, the  mixing parameter $|U|^2\equiv \sum_\alpha \theta_\alpha^2$ is also considered, characterizing the total interaction rate of neutrino singlets coupling to three lepton (neutrino and charged-lepton) flavours.  Since the explanation of the BAU via the TIRL mechanism favours hierarchical washout effects,  the total interaction rate of neutrino singlets will be dominated by a specific lepton flavour, yielding
\begin{align}\label{eq:Usq}
	\text{muon dominance}&: \quad |U|^2\equiv \sum_{\alpha}\theta_\alpha^2\approx \frac{|h_{\mu 1}|^2v^2}{M_1^2}+\frac{|h_{\mu 2}|^2v^2}{M_2^2}\,,
	\\[0.2cm]
	\text{electron dominance}&: \quad |U|^2\equiv \sum_{\alpha}\theta_\alpha^2\approx \frac{|h_{e1}|^2v^2}{M_1^2}+\frac{|h_{e 2}|^2v^2}{M_2^2}\,,
\end{align}
under the decoupling limit of $N_3$.

Each $(M_1,\theta_\alpha^2)$ plane
is shown as a slice at a fixed normalized mass difference $\Delta M/M_1$. Note that we do not impose any hierarchy between the two GeV-scale singlet neutrinos. This allows us to see if a definite mass hierarchy or mass degeneracy is preferred in TIRL, and also allows a more direct comparison with the  other two low-scale leptogenesis mechanisms, i.e., the ARS and RL, both of which rely on the mass degeneracy of the singlet neutrinos.

\begin{figure}[h]
	\centering
	\includegraphics[scale=0.345]{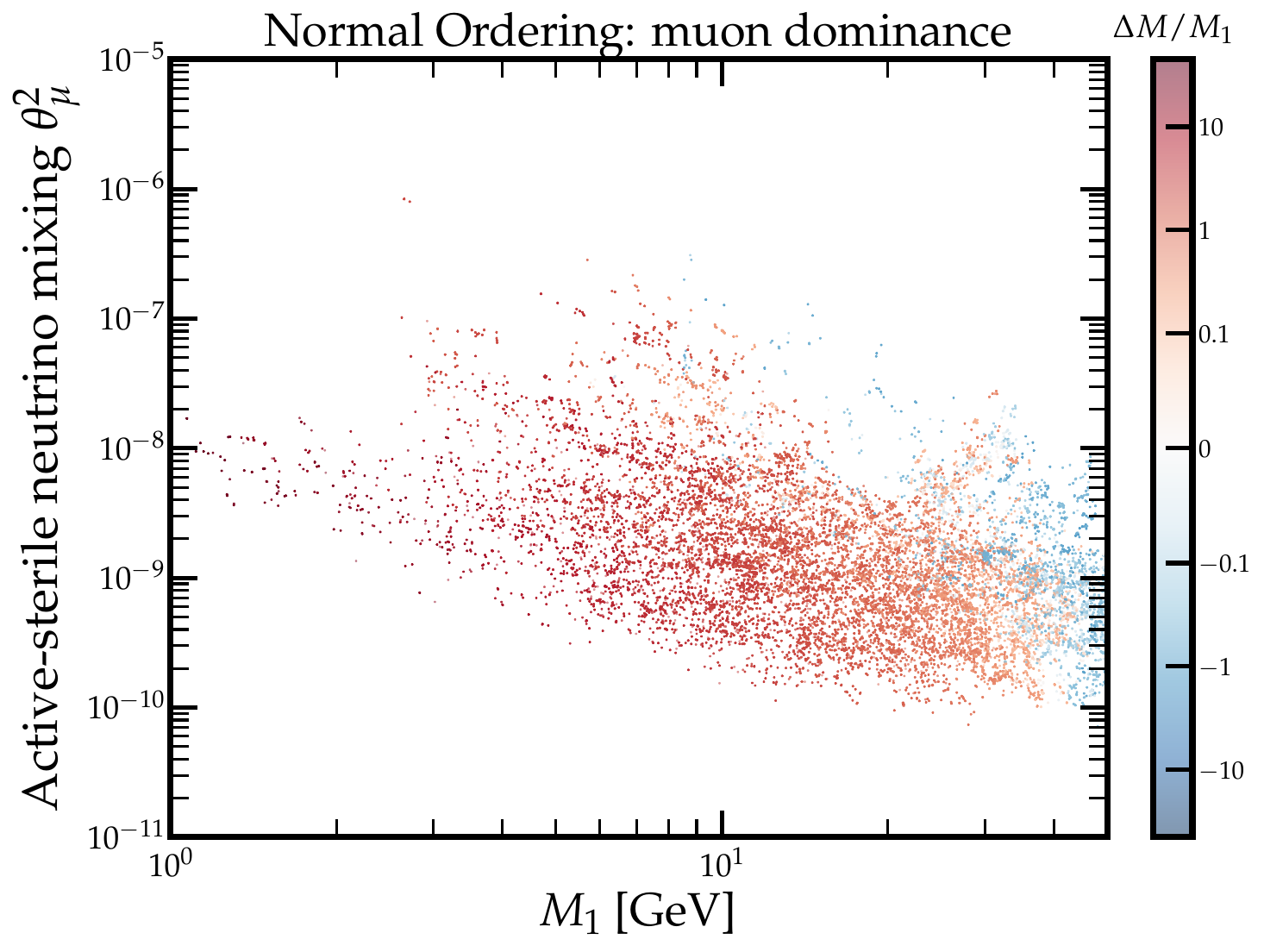} 
	\includegraphics[scale=0.345]{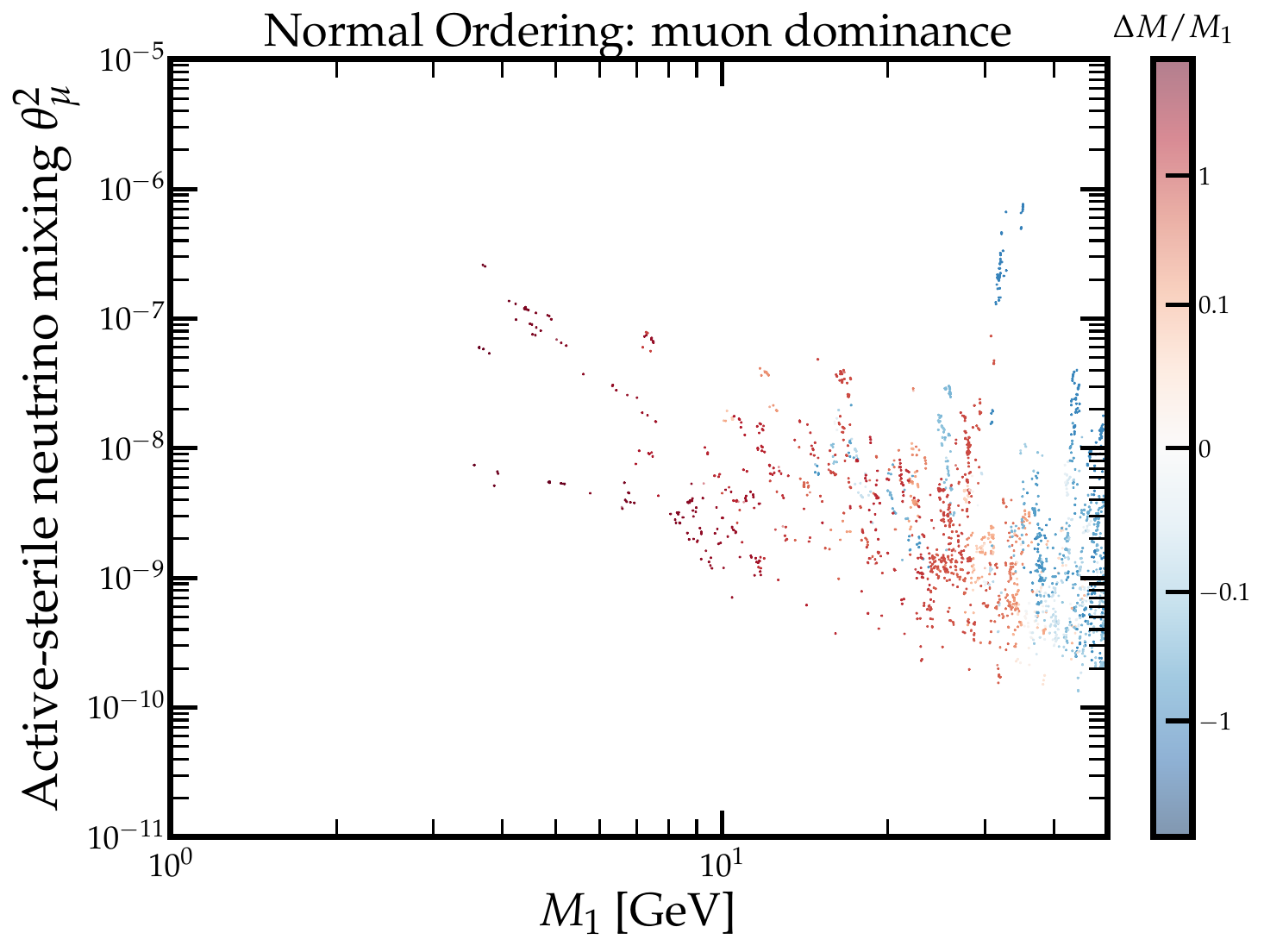}
	\\
	\includegraphics[scale=0.345]{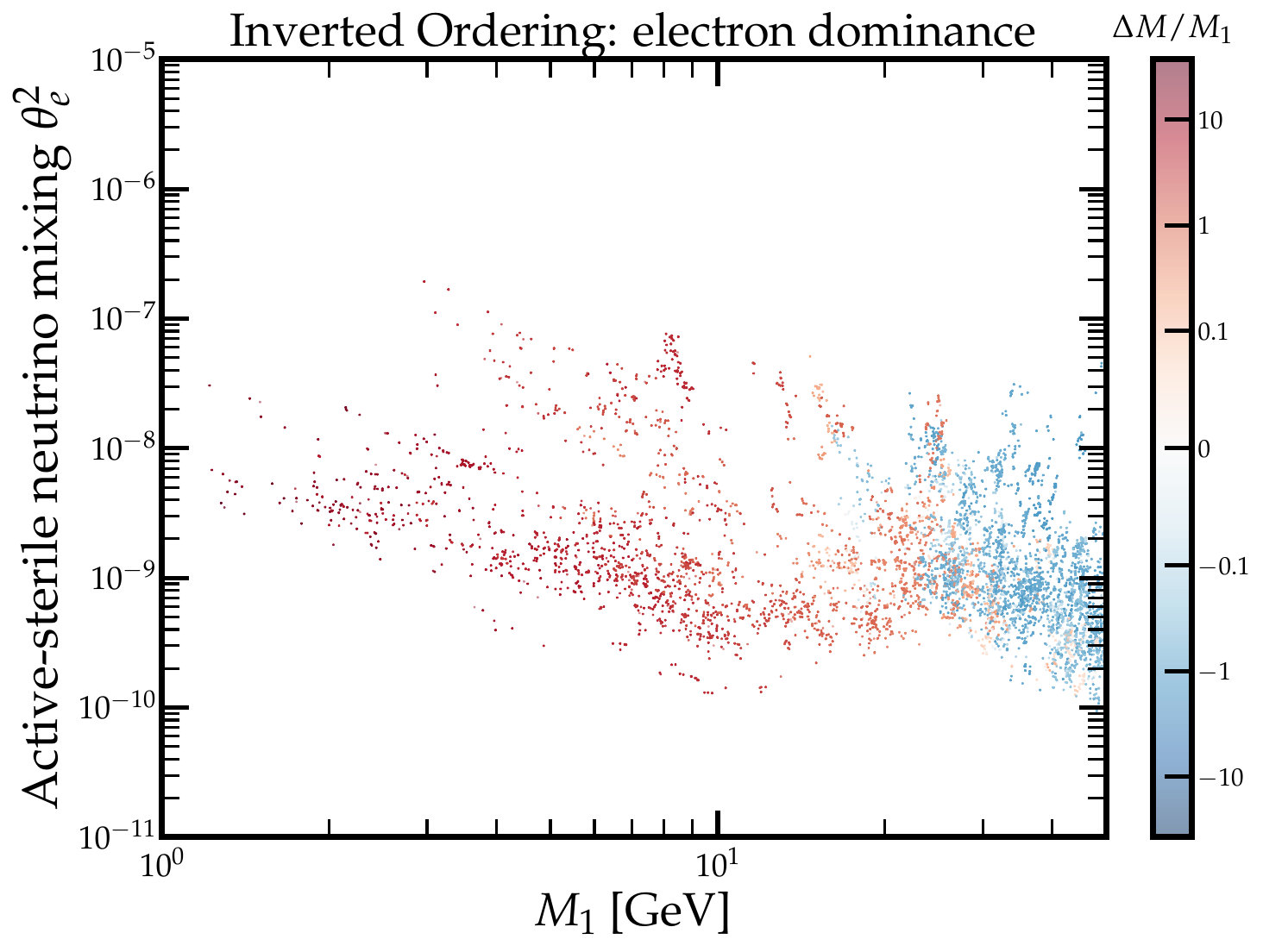} 
	\includegraphics[scale=0.345]{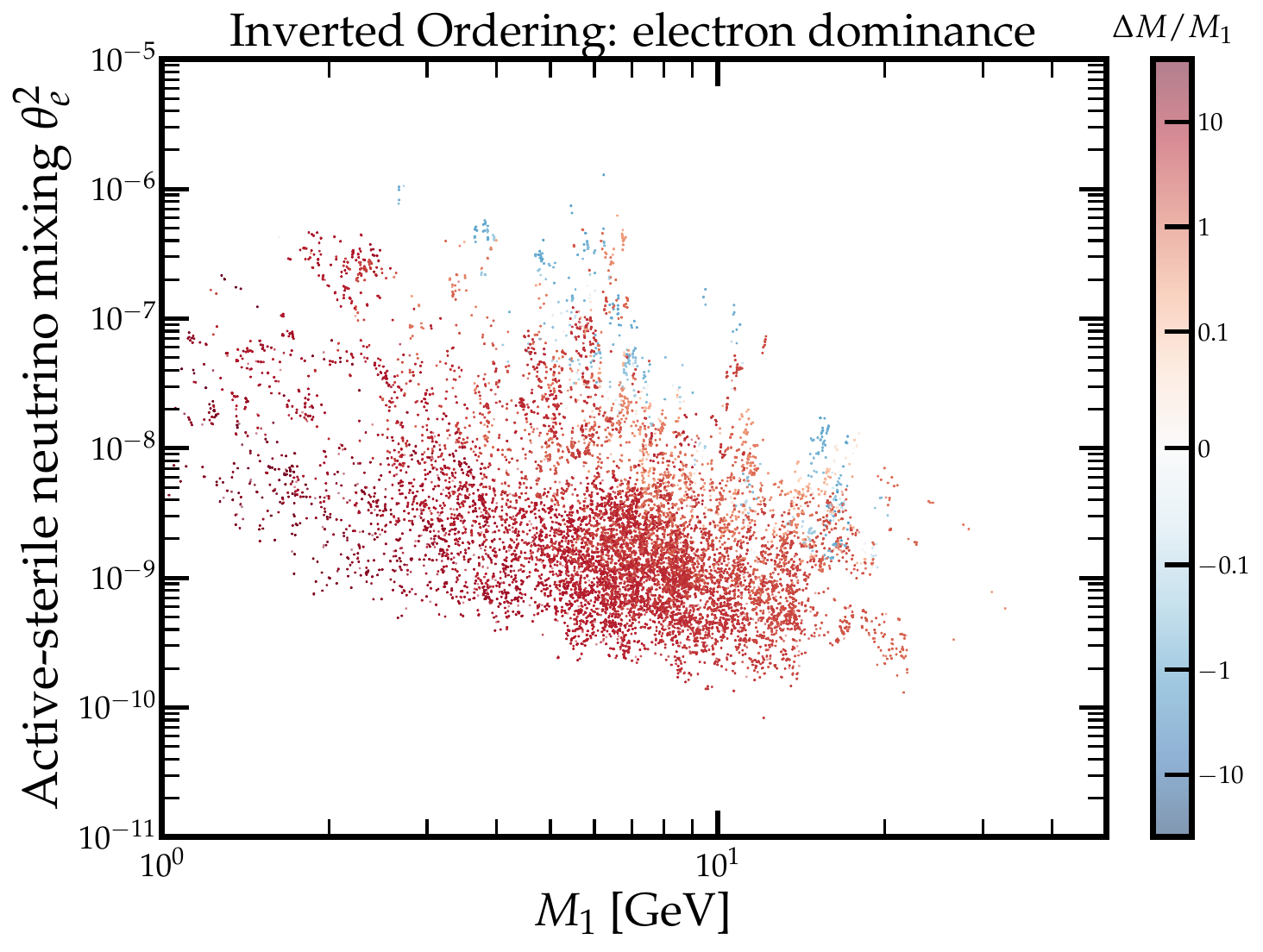}
	\caption{\label{fig:Mtheta_NOIO_MCMC} Scatter plots of the biased MCMC scan data that produce the correct order of $Y_B$ at $[10^{-11},10^{-10}]$. The upper panel shows the NO case, with the lightest SM neutrino mass fixed at $m_1=0.01$~eV (upper left) and $m_1=0$~eV (upper right). The lower panel shows the IO case, with the lightest SM neutrino mass fixed at $m_3=0.01$~eV (lower left) and $m_3=0$~eV (lower right). }
\end{figure}

\begin{figure}[t!]
	\centering
	\includegraphics[scale=0.345]{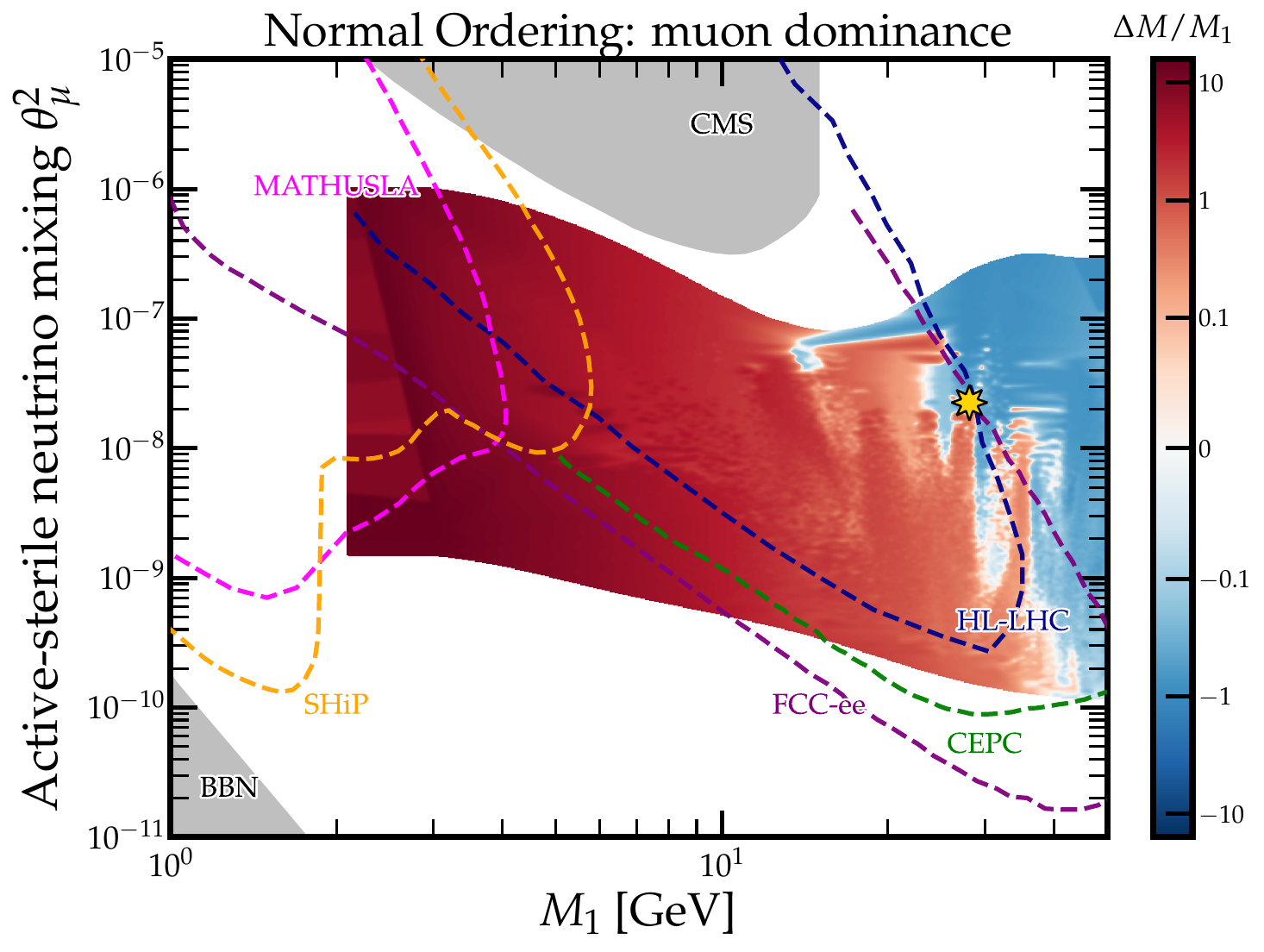} 
	\includegraphics[scale=0.34]{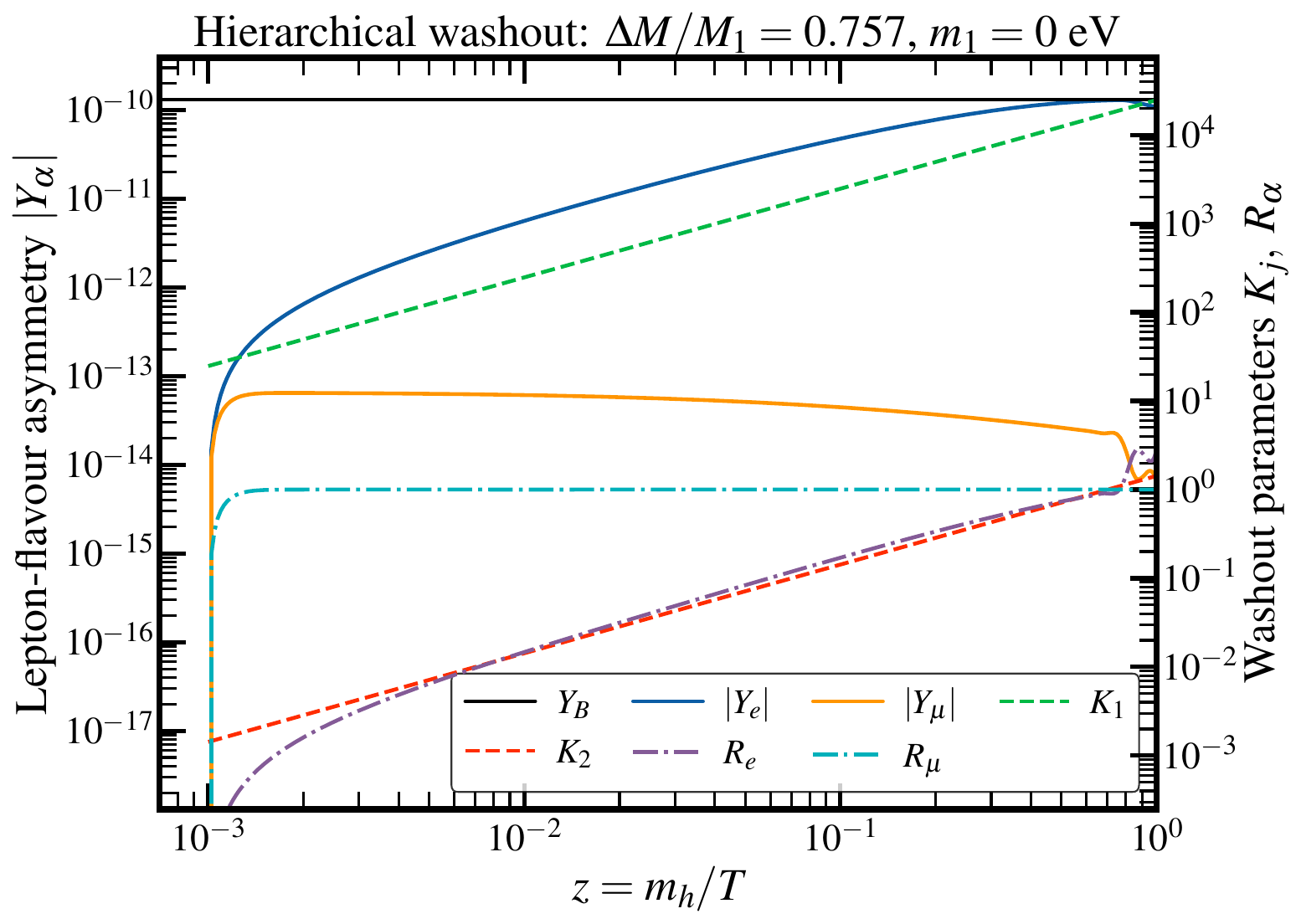}
	\\
	\includegraphics[scale=0.345]{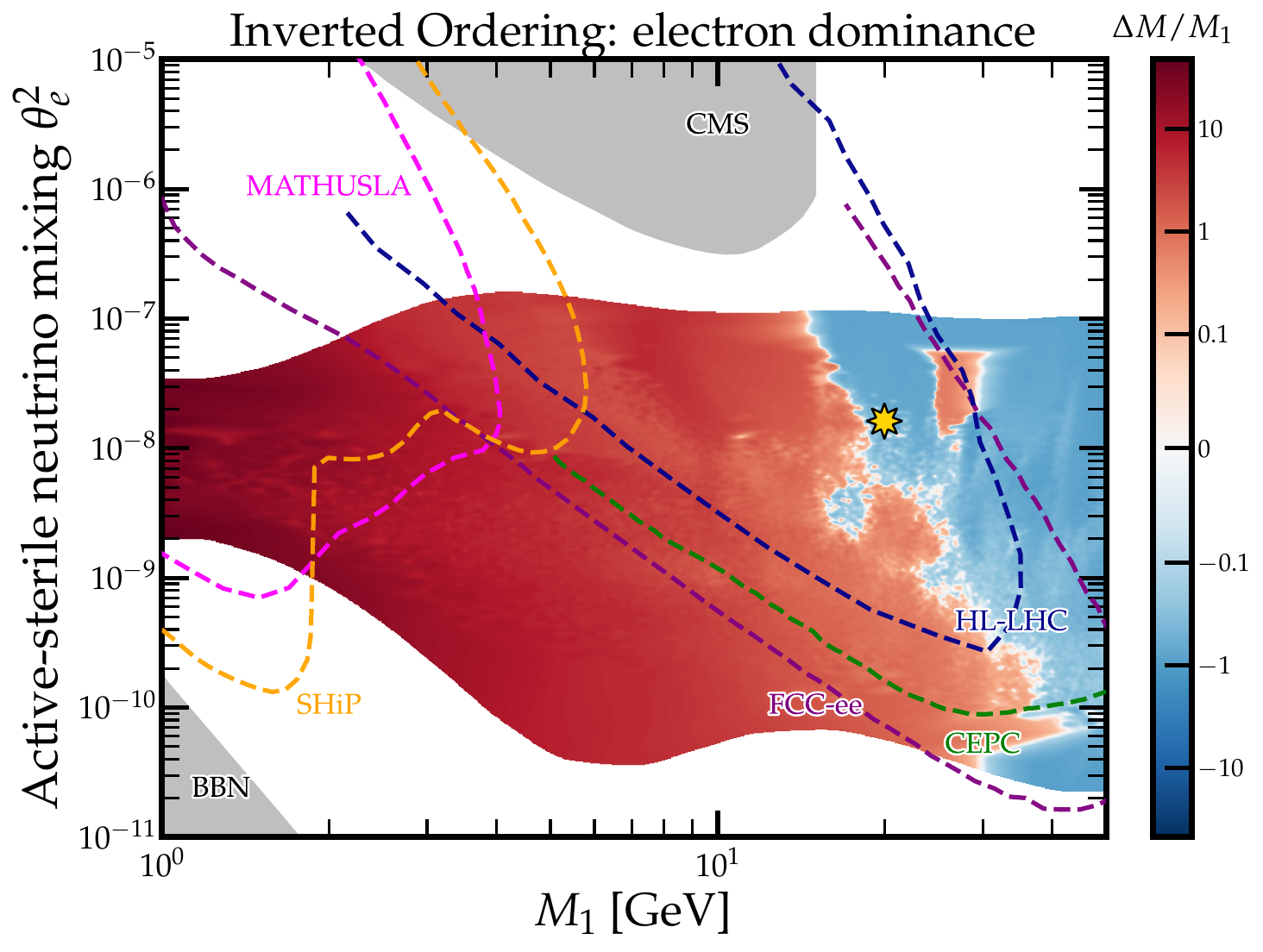}
	\includegraphics[scale=0.34]{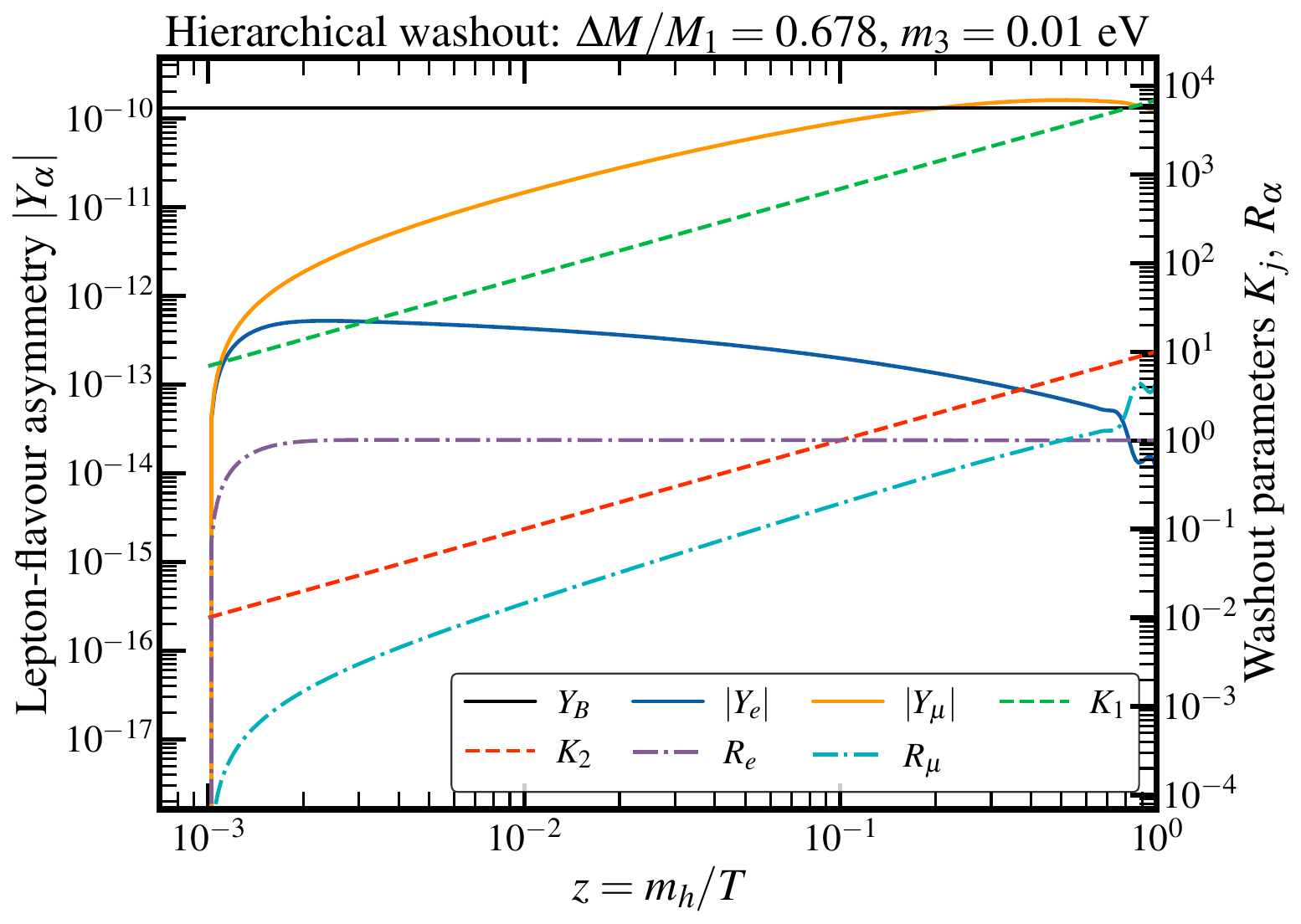}
	\\
	\includegraphics[scale=0.345]{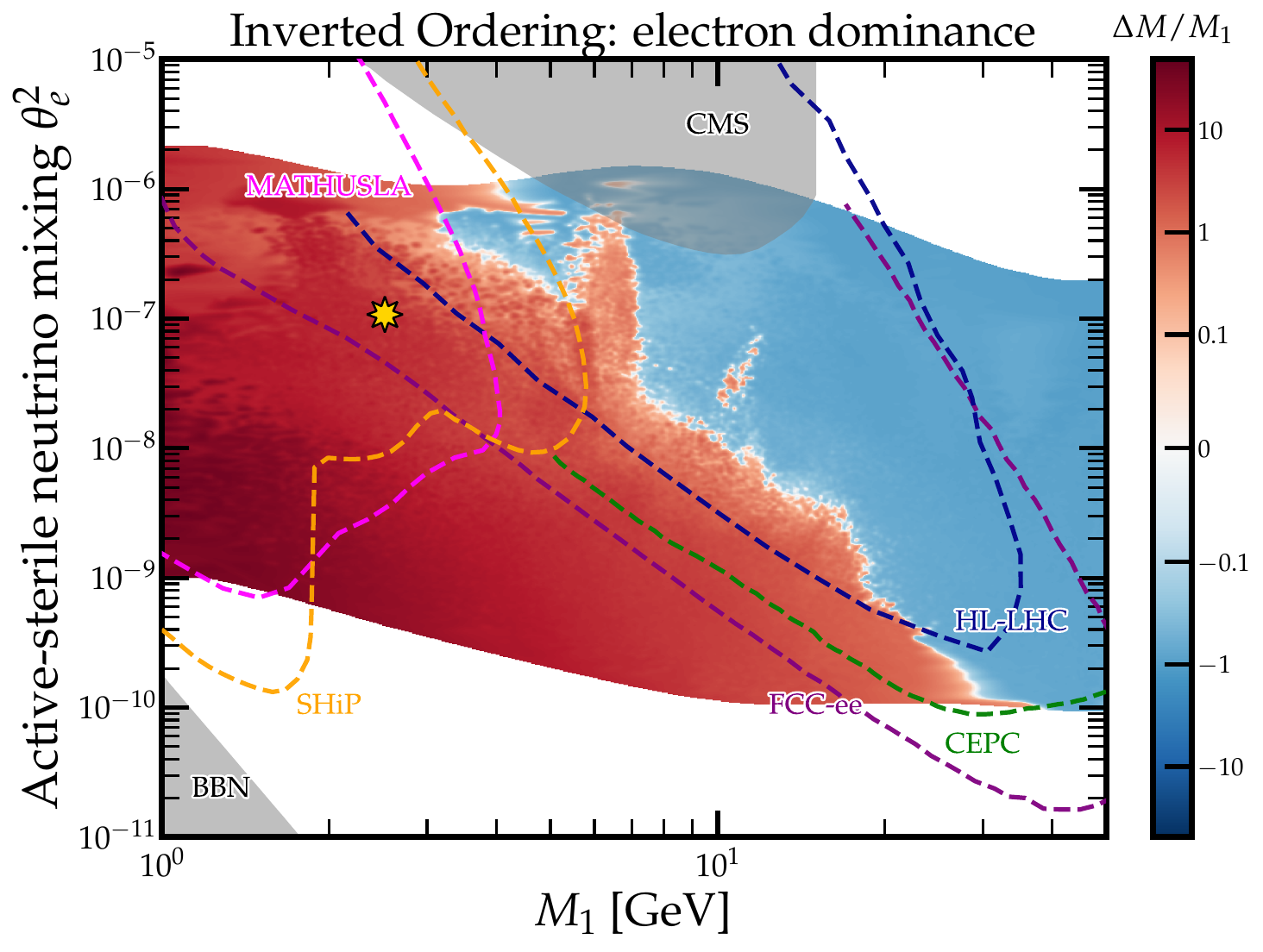}
	\includegraphics[scale=0.34]{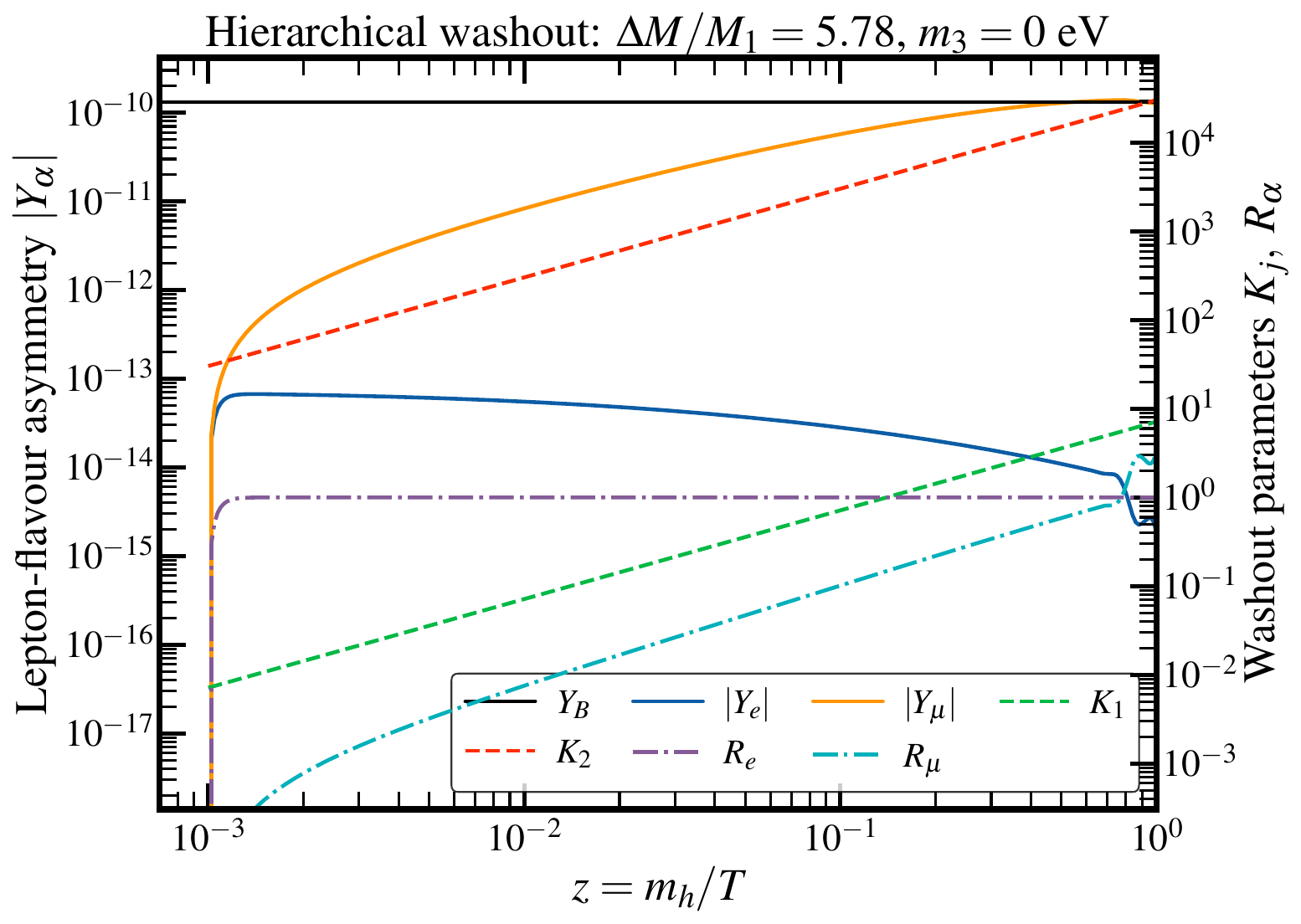}
	\caption{\label{fig:NOIO} Complementary numerical results to the one presented in the Letter. The left panels show general parameter space obtained from DNN training that  yields $Y_B=[10^{-11},10^{-10}]$, while the right panels correspond to the benchmark scenarios  from the yellow star given in the left panel.  \textit{Top panel}: the NO case with $m_1=0$. \textit{Middle panel}: the IO case with $m_3=0.01$~eV. \textit{Bottom panel}: the IO case with $m_3=0$.}
\end{figure}

Figure~\ref{fig:Mtheta_NOIO_MCMC} shows the scatter plots obtained from the biased MCMC scan.  While certain parameter space that produces the correct order of $Y_B$ (including the observed value) exists both in the NO and IO cases, the scatter points are fairly sparse and do not exhibit a clear dense clustering trend even under the biased MCMC scan with $10^6$ sample points. To obtain the general parameter space without missing the potentially feasible parameter points, which usually happens in multi-dimensional scan, we further train the data from Fig.~\ref{fig:Mtheta_NOIO_MCMC} by adopting the DNN normalizing flow.
Indeed, as shown in the left panels of Fig.~\ref{fig:NOIO},
the normalizing flow allows us densely resample and smoothly interpolate the viable parameter region for plotting, filling in gaps left by the finite, autocorrelated MCMC chain. The flow, serving as a cheap resampling and smoothing step after the hard MCMC exploration work, also avoids having to extend the expensive MCMC chain, since each further MCMC sample still needs a full kinetic-equation solver to get denser, less-autocorrelated coverage for plotting.

Figure~\ref{fig:NOIO} presents more numerical details about the general parameter space obtained from DNN training (left panels) and benchmark scenarios (right panels) complementary to the one shown in the Letter, confirming thereby the conclusions drawn in the main text.  In general, no mass degeneracy is required in TIRL to enhance the generation of $Y_B$, a distinctive feature from the ARS and RL mechanisms.  Overall, except for the different flavour washout,  the parameter space for the IO case  exhibits certain similar patterns to the NO case, up to some noticeable points we discuss below.  In the NO case with the lightest neutrino  being massless, there is a lower mass cut on $M_1$, where $M_1\lesssim 2$~GeV cannot generate the required order of $Y_B$. However, a vanishing $m_3$ in the IO case does not have such a mass cut. This can be understood by the CI parametrization. For Yukawa couplings $h_{\alpha i}\propto U_{\nu,\alpha\beta}\sqrt{m_\beta}\mathbb{R}_{\beta i}\sqrt{M_i}$ under the same order of rotation matrix elements $\mathbb{R}_{\beta i}$, reducing the SM neutrino mass
should be compensated for by a larger mass $M_i$ to keep $|h_{\alpha i}|^2$ in the range needed for successful TIRL. Although $m_1$ is always the smallest light
mass in NO, its contribution to the lepton asymmetry is not negligible in the muon dominance case: it relates to the PMNS matrix element $U_{\nu, \mu 1}$ which is not parametrically suppressed.  Using the
mass-squared difference for SM neutrinos, the contribution with $m_1=0.01$~eV comparing to the largest one reads 
\begin{align}
	\frac{|U_{\mu1}|\sqrt{m_1}}{|U_{\mu3}|\sqrt{m_3}}
	\approx 0.21\text{--}0.32\,,
\end{align}
implying that $m_1$ contributes roughly $20\%$--$30\%$ of the amplitude that $m_3$ does. This is large enough such that the low-$M_1$
configurations can rely on this extra contribution significantly to reach the correct-order of $Y_B$. Analogously, in the IO case with the electron dominance, the contribution with $m_3=0.01$~eV comparing to the largest one gives
\begin{align}
	\frac{|U_{e 3}|\sqrt{m_3}}{|U_{e 1}|\sqrt{m_1}}
	\approx 0.08\,,
\end{align}
which, owing to a much smaller $U_{e3}$ matrix element, can be too  small for the low-$M_1$
configurations to have a strong dependence on varying $m_3$. 

Another reason that allows significant parameter space with $m_3=0$ when comparing to what happens in NO between $m_1=0.01$~eV and $m_1=0$ comes from more flexible tuning among Yukawa couplings. Comparing the middle  and bottom panels of Fig.~\ref{fig:NOIO}, we see that the parameter space for the $m_3=0$~eV case turns to be certain broader towards larger Yukawa couplings. This could be a numerical coincidence arising from a stronger tuning among the Yukawa couplings, unless it is hinting at some underlying flavour symmetry.  Generally, a larger mixing angle $\theta_e$ is  a result of increasing $\text{Im}\omega_{ij}$ in the CI parametrization. Given that both the washout and CP-violating source will be enhanced, a net, enhanced lepton flavour asymmetry requires a more delicate cancellation to render the washout effect smaller. Indeed, we find  from the two corresponding benchmark scenarios shown in the right panels, that the $m_3=0$~eV scenario requires a much larger Yukawa cut $|h_e|/|h_\mu|=144$, nearly a factor of 3 larger than from the $m_3=0.01$~eV scenario. From the scan, we also noticed that allowing larger tuning to open the parameter space is more easily realizable in the IO case. Interestingly, similar situation also occurs in the ARS mechanism, where a  mass splitting as large as $\Delta M/M_N\sim 0.1$ between two GeV-scale sterile neutrinos was found to be feasible in the IO case~\cite{Klaric:2020phc,Klaric:2021cpi}, which has long been thought to be challenging~\cite{Pilaftsis:2004xx,Pilaftsis:2005rv,Canetti:2012kh,Shuve:2014zua,Canetti:2014dka}.  However, this requires a huge ratio of the Yukawa couplings between the two neutrino flavours. Explicitly, a  large mass splitting at  $\Delta M/M_N\sim 0.1$ requires~\cite{Klaric:2021cpi}
\begin{align}
	\text{ARS:}\quad\frac{\sum_\alpha|h_{\alpha 2}|^2}{\sum_\alpha|h_{\alpha 3}|^2}\sim 10^6\,,
\end{align}
corresponding to the IO with $m_3=0$.  Note that we used the  neutrino flavour labelling $N_2,N_3$ the same as that usually defined in ARS, which is different from our convention here. From the  right panels of Fig.~\ref{fig:NOIO}, we can extract the ratio
\begin{align}
	\text{TIRL:}\quad \frac{\sum_\alpha|h_{\alpha 2}|^2}{\sum_\alpha|h_{\alpha 1}|^2}\sim 10^3 \,,
\end{align}
from $K_2/K_1$ or $K_1/K_2$. The different ratios in the ARS and TIRL shown above indicate that a much stronger tuning is required in the Yukawa couplings if only two generations of heavy neutrinos are introduced in the framework.

The benchmark scenario attached to each DNN plot explains the evolution of different lepton flavour asymmetries in the hierarchical washout regime.  To characterize these washout patterns, we consider both the conventional parameter $K$ defined as the production rate over the Hubble parameter, and the ratio of the washout to the source term, namely,
\begin{align}
	K_j\equiv \frac{\Gamma_j}{H}\,,\quad R_{\alpha}\equiv \frac{|W_\alpha|}{|S_\alpha|}\,,
\end{align}
where $H$ is the Hubble parameter, and $\Gamma_j\approx 0.01 |h_{j}|^2 T$ is the thermal production rate of heavy neutrino flavour $j$ with $|h_j|^2\equiv \sum_\alpha |h_{\alpha j}|^2$ and the soft gauge-boson scattering effects included~\cite{Anisimov:2010gy,Besak:2012qm,Garbrecht:2013bia,Laine:2013lka}. The parameter $K_j$ measures if and when the heavy neutrino singlet becomes thermally populated in leptogenesis, while the ratio $R_\alpha$  signals when the washout dominates the evolution of the lepton asymmetry in flavour $\alpha$ and hence directly determines which flavour asymmetry survives (electron or muon dominance). The muon dominance observed in the NO case indicates that the washout effect in the  muon flavour asymmetry is much stronger than in the electron flavour, so that $Y_e$ eventually survives. This is clearly seen via $R_{\mu}\gg R_e$. The opposite appears in the IO case, where $Y_\mu$ eventually survives and $R_{\mu}\ll R_e$. For both cases, the survival of the flavour asymmetry $\alpha$ is guaranteed by the fact that $R_\alpha>1$ only shows up at the very end of leptogenesis $z=m_h/T_{\rm sph}\approx 0.95$, maximally reducing the washout effect. On the other hand, the large difference of the ratio $K_i/K_j$ for the two GeV-scale neutrino singlets further strengthens the accumulation of lepton flavour asymmetry. For instance, the hierarchy $K_1\gg 1\gg K_2$ indicates that $N_1$ will establish thermal equilibrium before sphaleron decoupling while $N_2$ will only become thermally populated at the very end of leptogenesis, allowing to a large extent for increasing the CP-violating source via larger Yukawa couplings.

For clarity, we document in order the Yukawa matrix elements $h_{\alpha i}$ and the masses of the GeV-scale neutrino singlets  for the four benchmark scenarios.
\begin{align}
	\text{NO}\,, m_1&=0.01~\text{eV}, M_1=8.03~\text{GeV}\,,M_2=40.25~\text{GeV}\,,
	\nonumber \\[0.2cm]
	h&= \begin{pmatrix}
		0.0167+0.0426i   & -0.0963-0.139i& (0.0218-0.0160i)\sqrt{M_3/\text{GeV}} \\
		0.0465-0.391i & -17.7+1.14i &  (-0.181-2.79i)\sqrt{M_3/\text{GeV}}  \\
		-0.0424-0.0126i &  -2.11-0.0891i &  (0.0152-0.332i)\sqrt{M_3/\text{GeV}} 
	\end{pmatrix}\times 10^{-6}\,,
	\\[0.3cm]
	\text{NO}\,, m_1&=0, M_1=28.12~\text{GeV}\,,M_2=49.39~\text{GeV}\,,
	\nonumber \\[0.2cm]
	h&= \begin{pmatrix}
		-0.117-0.189i   & 0.0464-0.0180i & (-0.0357+0.0220i)\sqrt{M_3/\text{GeV}} \\
		15.7-6.91i &0.146-0.0226i &  (-1.30-2.96i)\sqrt{M_3/\text{GeV}}  \\
		11.5-3.10i &  0.0185+0.0240i &  (-0.585-2.16i)\sqrt{M_3/\text{GeV}} 
	\end{pmatrix}\times 10^{-6}\,,
	\\[0.3cm]
	\text{IO}\,, m_3&=0.01~\text{eV}, M_1=19.95~\text{GeV}\,,M_2=33.47~\text{GeV}\,,
	\nonumber \\[0.2cm]
	h&= \begin{pmatrix}
		10.2+1.39i  & 0.0345+0.395i& (-0.312+2.29i)\sqrt{M_3/\text{GeV}} \\
		0.174+0.0637i & 0.108+0.0316i &  (-0.0148+0.0384i)\sqrt{M_3/\text{GeV}}  \\
		-1.74-3.39i & 0.0771+0.0124i&  (0.760-0.389i)\sqrt{M_3/\text{GeV}} 
	\end{pmatrix}\times 10^{-6}\,,
	\\[0.3cm]
	\text{IO}\,, m_3&=0, M_1=2.52~\text{GeV}\,,M_2=17.07~\text{GeV}\,,
	\nonumber \\[0.2cm]
	h&= \begin{pmatrix}
		0.337-0.0973i   &3.67+22.3i& (-5.40+0.888i)\sqrt{M_3/\text{GeV}} \\
		0.0374+0.00526i &-0.150-0.0323i &  (0.00859-0.0362)\sqrt{M_3/\text{GeV}}  \\
		0.0360-0.0185i &  0.469+4.70i &  (-1.14+0.113i)\sqrt{M_3/\text{GeV}} 
	\end{pmatrix}\times 10^{-6}\,,
\end{align}
where we put back explicitly the dependence on the decoupled $N_3$. We see that the third column  $h_{\alpha 3}$  in all these benchmark scenarios can be readily small for $M_3\ll M_1, M_2$. It becomes clear that for the setup of numerical scan by decoupling the third neutrino singlet,  the phenomenology related to $N_3$ can be  more predictable once its mass is known, since   the Yukawa couplings $h_{\alpha 3}$ is partly fixed by the BAU explanation.

\end{document}